\documentclass[runningheads]{llncs}
\usepackage[T1]{fontenc}
\usepackage{graphicx}
\usepackage{cite}
\usepackage{amsmath,amssymb,amsfonts}
\usepackage{graphicx}
\usepackage{textcomp}
\usepackage{xcolor}
\usepackage{tikz}
\usepackage[ruled, linesnumbered, noend]{algorithm2e}
\usepackage[table, dvipsnames]{xcolor}
\usepackage{wrapfig}
\usepackage{subcaption}
\usepackage{floatrow}
\usepackage[normalem]{ulem}

\usetikzlibrary{fit,shapes.misc}
\def\BibTeX{{\rm B\kern-.05em{\sc i\kern-.025em b}\kern-.08em
    T\kern-.1667em\lower.7ex\hbox{E}\kern-.125emX}}

\definecolor{MyResponseColor}{HTML}{2AA63E}

\definecolor{MyComment}{HTML}{F6339A}
\definecolor{myyellow}{HTML}{FFFAA0}
\definecolor{mygreen}{HTML}{C1E1C1}
\makeatletter
\DeclareRobustCommand*\cal{\@fontswitch\relax\mathcal}
\makeatother

\newcommand{\kw}[1]{{\ensuremath {\mathsf{#1}}}\xspace}

\newcommand{\stitle}[1]{\noindent{\bf #1}}

\renewcommand{\paragraph}[1]{\vspace{2mm}\noindent {\bf #1:}}

\newcommand{\eat}[1]{}
\newcommand{\crminer}{\kw{CR}-\kw{Miner}}

\makeatletter
\DeclareRobustCommand*\cal{\@fontswitch\relax\mathcal}
\makeatother

\newcommand{\sstab}{\rule{0pt}{8pt}\\[-1.8ex]}

\newcommand{\bi}{\begin{itemize}}
	\newcommand{\ei}{\end{itemize}}
	{\end{itemize}} %\vspace{-0.5ex}}

\begin{document}

\title{Data Profiling for Change Rules}

\author{Nishttha Sharma\orcidID{0009-0009-0729-0761} \and
Fei Chiang\orcidID{0000-0003-4128-8074} }
\authorrunning{F. Author et al.}
\institute{Dept. of Computing and Software \\ 
McMaster University, Hamilton, ON, Canada \\
\email{\{sharmn99, fchiang\}@mcmaster.ca}}
\authorrunning{N. Sharma and F. Chiang}

% \section*{Response Letter}
%     \input{Chapters/ResponseDoc}

\newpage
\maketitle              % typeset the header of the contribution
%
% \vspace{-0.5cm}

\begin{abstract}
Understanding data change is critical towards understanding trends, normal vs. abnormal behaviours, recognizing patterns, and the causes of change. Existing database systems have limited support for change management, relying on statistics, triggers, and constraints. Data quality rules model sequential changes along a restricted set of attributes, quantify change among unordered tuples, and have limited ability to model the context under which attribute changes occur. In this paper, we introduce \emph{Change Rules (CRs)} that quantify the sequential changes among ordered tuples in both the antecedent and consequent attributes. CRs aim to address the limitations of existing declarative dependencies to support trend analysis and causal relationships that trigger change among attributes. We propose \crminer, an automated algorithm for CR discovery that generates candidate change intervals in a level-wise manner. Experimental results show that \crminer achieves an average runtime improvement of 40--50\% over existing baselines. 
%a state-of-the-art dependency discovery method.

%We study the problem of discovering dependencies between attribute changes in ordered relational data. To address this, we propose a framework that models attribute-wise changes using differential functions over consecutive tuples and encodes them using bitset representations. Based on these, we introduce a level-wise candidate generation and pruning strategy to identify minimal and continuous change intervals. We further propose a rule discovery mechanism that extracts change dependencies by analyzing overlapping continuous segments across attributes. Experiments on real-world datasets show that our approach is scalable with respect to dataset size and number of attributes, while consistently discovering more meaningful dependencies under strong support constraints compared to baseline methods.
% \vspace{-0.2cm}

% \vspace{-0.5cm}
\end{abstract}

\section{Introduction}
    % \vspace{-0.5cm}

Data analysts seek to understand the cause and impact of change: What changes occurred (in the last day)? What caused these changes? Do similar update patterns exist? Was there an event that triggered a large set of changes, and what was the impact? How long have these changes been occurring? How can we minimize the impact of an undesirable change? Organizations have little intuition and knowledge of the types of data changes occurring in their systems to answer these questions. Existing database technology provides extensive features to efficiently retrieve and analyze data. However, database functionality to support change management is limited to vendor products that mine transaction logs, triggers, and check constraints that react to specific value changes, statistics that are re-generated for query optimization, and explanations for query execution plans. There is minimal work towards exploring, identifying, and analyzing change in data, and its impact on data quality.  This lack of tools for interpreting temporal and spatial changes hinders our ability to completely understand the evolution of an entity, its evolving relationships to other entities, and trends. 

Real-world data \eat{rarely remains static as data}continuously changes over time. This evolution introduces challenges in understanding and managing changes effectively. These changes often carry critical information, revealing patterns and trends \eat{, and triggers}that are essential for understanding environmental conditions, system, and user behaviour\eat{ and trends}. From a data quality perspective, identifying unexpected changes is critical to recognizing malicious activity and improper data use. Declarative methods have proposed data quality rules such as Order Dependencies (ODs) that capture relative order among tuples according to a set of attributes (e.g., RBC $\mapsto$ Hemoglobin), but do not quantify the magnitude of change~\cite{szlichta2016effective}. Sequential Dependencies (SDs) constrain the amount of change in a consequent attribute across ordered tuples (along the antecedent attribute) but do not quantify the change in the antecedent attributes~\cite{golab2009sequential}. Differential Dependencies (DDs) capture bounded differences between tuple pairs along a set of attributes, but are defined over \textit{unordered} tuples, and do not model changes between ordered tuples~\cite{song2011differential}. Consider the following example highlighting limitations of existing data quality rules.

\begin{example}
Table~\ref{tab:example} shows a sample of real patient blood lab results collected at various time points from the MIMIC-III \cite{johnson2016mimic}.  The records represent Red Blood Cell (RBC), Hemoglobin (Hbg), Hematocrit (Hct), White Blood Cell (WBC), and Platelet counts.  Table~\ref{tab:dep} shows defined OD, SD and DD rules over this data. The OD states that for tuples ordered (in ascending order) on RBC, the values in Hbg should also be ascending. There are no specifications on the consecutive differences between tuple values in the antecedent nor consequent attributes. The SD states that for tuples ordered on RBC, sequential changes in Hbg should occur in the range of (0, 1), but does not define the ranges in which antecedent attribute changes (RBC) should occur.  Lastly, the DD states that if the pairwise difference among all tuple pairs in RBC is less than 1.2, then the difference between the same tuple pairs in Hbg should be less than 3.5. However, these bounds are defined for non-ordered tuple pairs, making them inapplicable to identify sequential changes and trends \emph{in both antecedent and consequent attributes}.
\end{example}

\begin{table}[t]
    \centering
    \begin{floatrow}
        
        % --- FIRST TABLE ---
        \ttabbox{%
            \caption{Patient lab results}%
            \label{tab:example}%
        }{%
            \begin{tabular}{|c|c|c|c|c|c|}
            \hline
            & \textbf{RBC} & \textbf{Hbg} & \textbf{Hct} & \textbf{WBC} & \textbf{Platelets} \\ \hline
            $r_1$ & 3.02 & 9.7  & 28.2 & 6.2  & 80  \\ \hline
            $r_2$ & 3.2  & 10.1 & 28.6 & 11.3 & 108 \\ \hline
            $r_3$ & 3.2  & 10.1 & 28.8 & 13.4 & 107 \\ \hline
            \rowcolor{myyellow}$r_4$ & 3.52 & 10.7 & 31.8 & 8.7  & 108 \\ \hline
            \rowcolor{myyellow}$r_5$ & 3.72 & 11.4 & 29.4 & 10.9 & 103 \\ \hline
            $r_6$ & 3.82 & 11.8 & 30.9 & 13.4 & 129 \\ \hline
            $r_7$ & 4.01 & 12   & 36.2 & 11.2 & 123 \\ \hline
            $r_8$ & 4.13 & 12.9 & 38   & 12.1 & 141 \\ \hline
            \end{tabular}%
        }

        % --- SECOND TABLE ---
        \ttabbox{%
            \caption{Table~\ref{tab:example} Dependencies.}%
            \label{tab:dep}%
        }{%
            \begin{tabular}{|c|c|}
            \hline
            \textbf{Type} & \textbf{Dependency} \\
            \hline
            OD & $\text{RBC} \mapsto \text{Hbg}$ \\
            SD & $\text{RBC} \rightarrow_{(0,1)} \text{Hbg}$ \\
            DD & $\text{RBC}(\le1.2) \rightarrow \text{Hbg}(\le3.5)$ \\
            CR & $\text{RBC}\;\;_{(0,2)} \rightarrow_{(0.2,0.9)} \text{Hbg}$ \\
            \hline
            \end{tabular}%
        }

    \end{floatrow}
\end{table}

We address these limitations by introducing a new type of rule called \emph{Change Rules (CRs)} that define bounded changes in both the antecedent and consequent attributes.  For example, a CR such as the one shown in Table~\ref{tab:dep}, specifies that if the difference between consecutive RBC attribute values occurs between (0, 2), then the corresponding difference in Hbg should be in the range (0.2, 0.9).  

However, not all observed changes necessarily contribute equally when determining these bounds. For example, the highlighted tuples $(r_4,r_5)$ show a small increase in RBC and Hbg but a decline in Hct. Such transitions introduce irregular variation in Hct despite stable trends in closely related attributes. \eat{, and if treated equally, would expand the range of valid Hct changes associated with RBC.} To determine whether a data change is normal (or not), it is necessary to consider related attributes that provide context of the change.  We study how to compute a scaling factor that adjusts data changes for this necessary context.

\noindent \textbf{Challenges.} Our work aims to identify relevant changes that are contextualized with related attributes, and to develop an algorithm for change rule discovery. We address the following challenges:  \\
(1) Identifying relevant context for attribute changes:  changes among attribute values are not always proportional and occur within a context of related attributes. For example, a larger salary increase in a given year may be due to an increase in the number of employees, lower spending costs, and increased sales revenue. Identifying relevant attributes and their changes, particularly with a large number of attributes and records, requires efficient, scalable solutions. \\
(2) Manual definition of CRs is \eat{laborious and }time-consuming and requires domain expertise. Mining CRs is challenging for growing schema and data size, as we must consider the scalability of rule discovery to provide effective pruning strategies. \eat{to ensure computational efficiency.} \\
(3) Given the large number of possible CRs, \eat{that are possible,}identifying meaningful and relevant CRs requires a notion of \emph{minimality}. Intuitively, minimal CRs should have the fewest number of antecedent and consequent attributes, and the shortest interval for each attribute, while satisfying minimum frequency (support) thresholds.

\noindent \textbf{Contributions.}  We make the following contributions: 

% \vspace{-0.2cm}

\sstab 
(1) \uline{Introduce and formally define \emph{Change Rules (CRs)}} that extend existing data dependencies to consider bounded changes in the antecedent and consequent attributes among an ordered set of tuples. 

% \vspace{-0.2cm}

\sstab 
(2) \uline{\crminer: a new sequential algorithm for CR discovery.}  \crminer efficiently generates differences among tuples (called differential functions), and uses these functions to compute minimal width intervals that maximize the number of satisfying pairwise tuples, satisfying a predefined support threshold.  We leverage bitset representations for support computation via longest consecutive subsequences for improved algorithmic efficiency.

% \vspace{-0.2cm}

\sstab 
(3) We propose a methodology to \uline{contextualize changes} that adjust changes in a target attribute using a learned context factor $\beta$. This contextualization provides explainability for `out-of-range' changes that may appear abnormal, but with appropriate context, improves explainability and ease of interpretation. 

% \vspace{-0.2cm}

\sstab 
(4) \uline{Experimental system evaluation.} We evaluate \crminer\ on three real-world datasets, MIMIC-III, Employment, and Weather, to validate its scalability and effectiveness. We show that \crminer\ significantly outperforms existing baselines, achieving a 40-80\% speedup over FastDD on the MIMIC-III and Employment datasets, and a 50-85\% speedup on the Weather dataset.

\section{Preliminaries}
    We define the necessary definitions and notation. 

\noindent \textbf{Data model.} Consider a relational schema $R = \{A_1, A_2, \ldots, A_N\}$, and an instance $I = \langle r_1, r_2, \ldots, r_m \rangle$, where each tuple $r_k$ assigns a value $r_k[A_i]$ to attribute $A_i \in R$. We impose an ordering on $I$ based on attribute values. For a given attribute $A_i \in R$, the tuples are sorted in ascending order of $A_i$, producing an ordered sequence $\pi_i(I) = \langle r_1, r_2, \ldots, r_m\rangle$ such that $r_1[A_i] \le r_2[A_i] \le \ldots \le r_m[A_i]$. This ordering captures how other attributes evolve as the value of $A_i$ increases.

\noindent \textbf{Data change.} Given an ordered sequence $\pi_i(I) = \langle r_1, r_2, \ldots, r_m\rangle$, we define the change in $A_j \in R$ between two consecutive tuples $r_k$ and $r_{k+1}$ as:
% \[
% \Delta A = d(r_k[A], r_{k+1}[A])
% \]
\[
\delta_{(k,k+1)}[A_j] = d(r_k[A_j], r_{k+1}[A_j])
\]
where $d(\cdot)$ is a distance function that quantifies the change in $A_j$ \eat{has changed }between the two tuples. \eat{For numeric attributes, the distance function is defined as the difference between consecutive values, i.e., $d(r_k[A_j], r_{k+1}[A_j]) = r_{k+1}[A_j] - r_k[A_j]$. For non-numeric attributes, we use the Levenshtein distance to measure dissimilarity.}
Each consecutive tuple pair $(r_k, r_{k+1})$ is represented as a \textbf{change vector} as follows: $\overrightarrow{r_k} = \big( \delta_{(k,k+1)}[A_1], \delta_{(k,k+1)}[A_2], \ldots, \delta_{(k,k+1)}[A_N] \big)$.

\noindent \textbf{Differential functions.}  Differential functions impose constraints on the difference between attribute values across tuple pairs ~\cite{song2011differential}. A differential function $\lambda(A_j)$ over an attribute $A_j$ is of the form: $\lambda(A_j) = (\delta_{(k,k+1)} [A_j] \in [l,u])$, where $l$ and $u$ denote the lower and upper bounds on the change in $A_j$\eat{, respectively}. The function $\lambda(A_j)$ is satisfied for a tuple pair $(r_k, r_{k+1})$ if the change in $A_j$ lies within the interval. Let $\Psi$ denote the set of all differential functions across all attributes. 

\noindent \textbf{Diff-sets.} Diff-sets were introduced to efficiently encode violations of differential functions \cite{kuang2024efficient}. The diff-set for a tuple pair is the set of all differential functions in $\Psi$ that are not satisfied by the change $\delta_{(k,k+1)}[A_j]$ ($\forall A_j \in R$). Formally, the diff-set of a tuple pair is $D(r_k, r_{k+1}) = \{ \lambda(A_j) \in \Psi \mid \delta_{(k,k+1)} [A_j] \notin [l, u]\}$. The set of all such diff-sets is denoted as $D$. For a specific function $\lambda(A_j)$, we define  $D(\lambda(A_j)) = \{ U \in D \mid \lambda(A_j) \in U \}$, which represents the set of all tuple pairs that violate $\lambda(A_i)$.

% \vspace{-0.2cm}

\section{Framework Overview}
    \begin{figure*}[t]
    \centering
    \includegraphics[width=\linewidth]{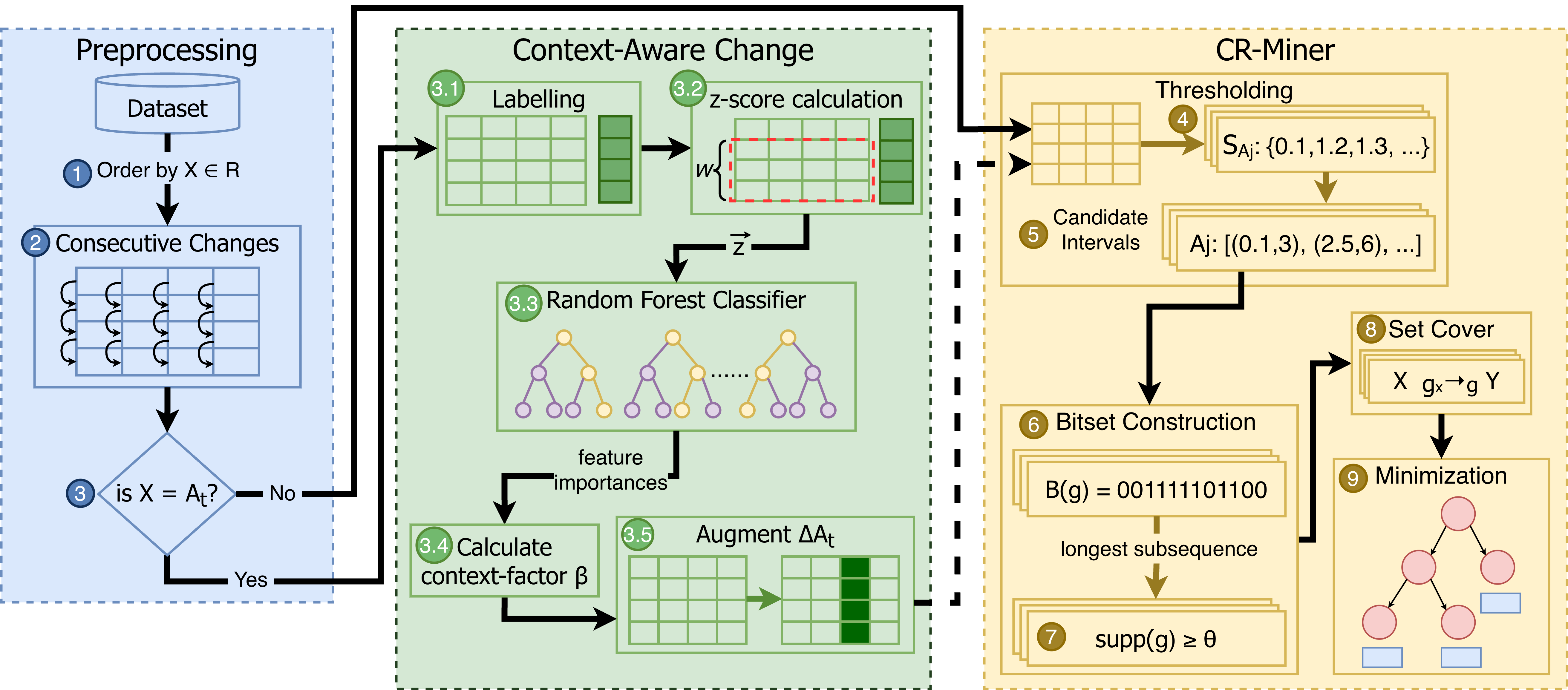}
    \caption{Framework Overview}
    \label{fig:framework}
\end{figure*}

% \vspace{-0.5cm}

Figure~\ref{fig:framework} gives an overview of the framework \eat{context-aware change identification and \crminer}, consisting of three phases: Preprocessing, Context-Aware Change, and \crminer.  In preprocessing, we order $I$ by $X$, and compute consecutive changes for each attribute $A_j \in R$ as $\delta_{(k,k+1)}[A_j] = d(r_k[A_j], r_{k+1}[A_j])$. 

\noindent \textbf{Context-aware changes.} For a given target attribute $A_t$, we seek to identify changes that lie outside of an expected range. We do this by defining a window of size $\mathcal{W}$ and computing the z-score of each $\delta_{(k,k+1)}[A_j], \forall A_j \in R$ producing a change vector $\overrightarrow{z_k} = (z_k[A_1], z_k[A_2], \ldots, z_k[A_N])$. Using statistical $z$-score cutoffs, we compute normal vs. abnormal labels with respect to the window $\mathcal{W}$ of changes. These vectors are the input to a Random Forest classifier that is used to identify whether changes in $A_t$ are normal or not, and the relative feature weights among the attributes $A_j \in R\setminus{A_t}$. We \eat{select the top-$p$ highest weight features to }use these weights to compute a context factor $\beta$ for each tuple pair of changes that adjusts the changes in $A_t$. \eat{ according to the most influential attribute changes in $A_j$.}  

%These normalized vectors are then labeled and used to train a Random Forest classifier that predicts whether the change in the target attribute is abnormal. Feature importance scores, computed using Mean Decrease Accuracy (MDA), quantify the contribution of each attribute’s change to the prediction. The top-$p$ most relevant attributes are selected to compute a context factor $\beta$ for each tuple pair. This factor is used to adjust the raw change in the target attribute, resulting in a context-aware change value.

% After this, a Random Forest Classifier is trained using these normalized change vectors and given labels to predict whether the change in the selected target attribute is abnormal or unexpected. Feature importance calculated by the classifier, using Mean Decrease Accuracy (MDA), indicates how much each feature (z-score of the change in an attribute $A_j$) contributes to the model’s prediction accuracy. We then use these feature importance scores to identify the top-$p$ most relevant attributes, which are used to compute a context factor $\beta$ for each tuple pair's $\delta_{(k,k+1)} [A_t]$. This factor, when multiplied to the raw change in the target attribute, adjusts it to produce a context-aware change value.

\noindent \textbf{\crminer.} 
For each attribute $A_j \in R$, we define the set of sorted changes as $S_{A_j}$. Candidate intervals are generated from $S_{A_j}$ by computing a bitset for each interval where the bit is set to 1 if the corresponding change lies in the given interval, and 0 otherwise. If the longest consecutive sequence of $1$'s for an interval gap $g$ satisfies the minimum segment coverage $\theta_c$, $g$ is retained as a valid differential function. Using these differential functions, we generate candidate CRs of the form $X\;\; _{g_x} \rightarrow_g \;\; Y$ via a set cover procedure over the corresponding bitsets of $g_x$ and $g$ whose overlap satisfies a minimum support threshold $\theta$. We seek minimal, non-redundant rules where the antecedent and consequent intervals cannot be further narrowed without reducing the support of the rule.

\section{Context-Aware Changes}

\begin{figure}[t]
    \centering
    \begin{subfigure}[b]{0.5\linewidth}
        \centering
        \includegraphics[width=0.8\linewidth]{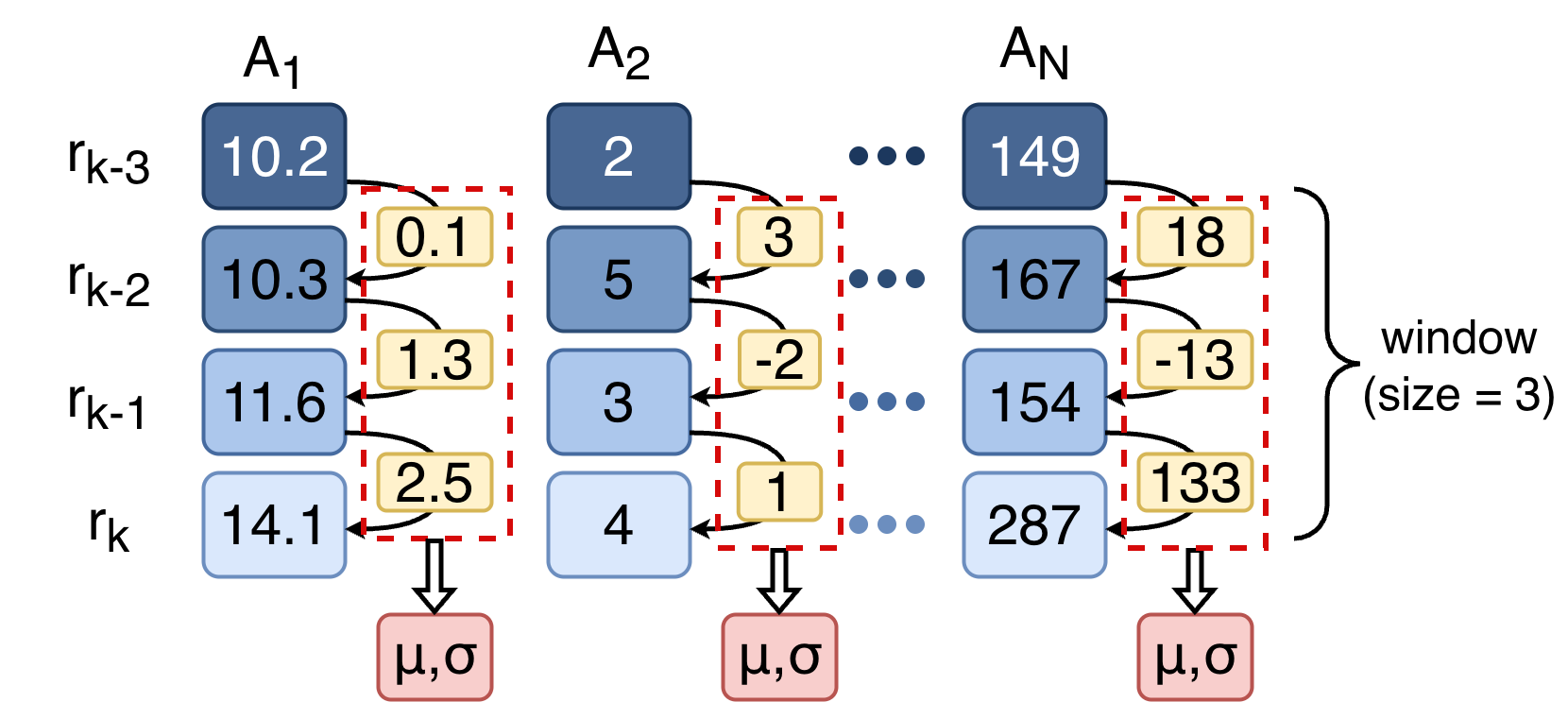}
        \caption{Localized changes within window $\mathcal{W}$}
        \label{fig:algo1}
    \end{subfigure}
    \hfill
    \begin{subfigure}[b]{0.45\linewidth}
        \centering
        \includegraphics[width=0.8\linewidth]{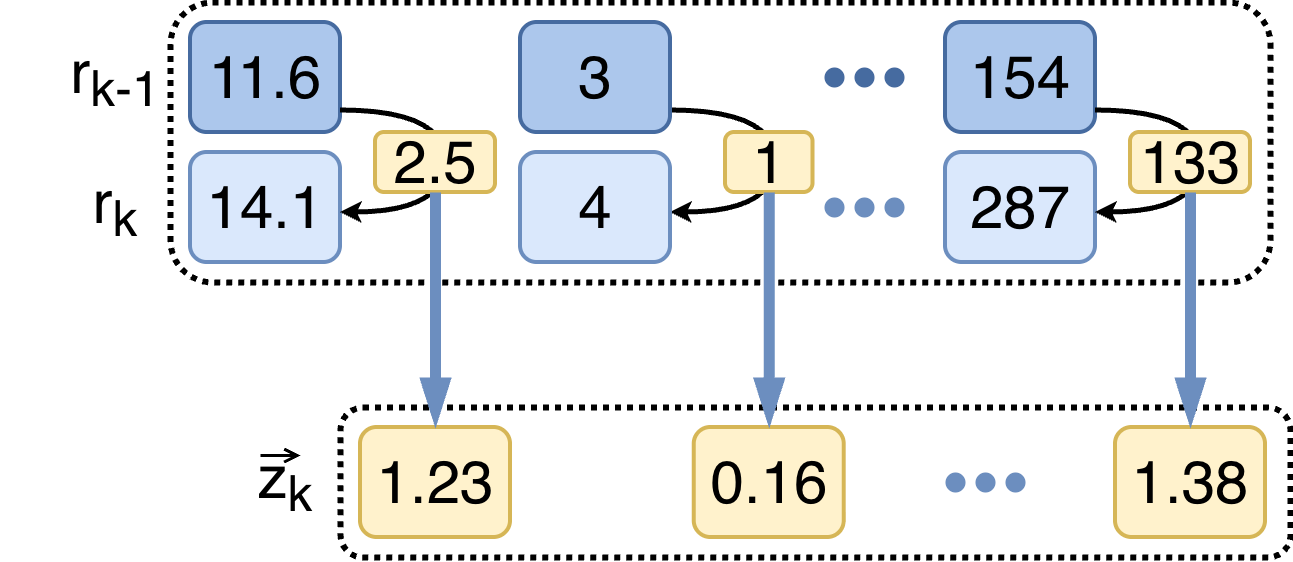}
        \caption{Normalized changes}
        \label{fig:algo2}
    \end{subfigure}
    \caption{Computing context-aware changes}
\end{figure}

Given an ordered sequence $\pi_t(I) = \langle r_1, r_2, \ldots, r_m \rangle$, each tuple pair $(r_k, r_{k+1})$ is associated with a change vector $\overrightarrow{r_k}$. For each attribute $A_j \in R$, we construct a window of the previous $\mathcal{W}$ change values $\{\delta_{(k-w,k-w+1)}[A_j], \ldots, \delta_{(k-1,k)}[A_j]\}$. From this window, we compute the mean $\mu_{k}[A_j]$ and standard deviation $\sigma_{k}[A_j]$, as shown in Figure~\ref{fig:algo1}. To make changes comparable across attributes with different scales, we apply z-normalization: $z_k[A_j] = \frac{\delta_{(k,k+1)}[A_j] - \mu_k[A_j]}{\sigma_k[A_j] + \epsilon}$, where $\epsilon$ is a small constant for numerical stability. Each tuple pair $(r_k, r_{k+1})$ is thus represented as a normalized change vector $\overrightarrow{z_k} = (z_k[A_1], z_k[A_2], \ldots, z_k[A_N])$, as shown in Figure~\ref{fig:algo2}.

\noindent \textbf{Learning weights.} For a chosen target attribute $A_t$, we train a classifier using $\overrightarrow{z_k}$ as input and labels indicating whether the observed change $\delta_{(k,k+1)}[A_t]$ is normal or not.  A change is marked abnormal if $z_k[A_t]$ exceeds a fixed threshold or significantly deviates from contextual behaviour. Formally, for each tuple pair, we compute a context baseline using the average z-score of contextual attributes, and assign a positive label if $z_k[A_t] > \eta$ or $z_k[A_t] > \nu \cdot avg(\text{context})$, where $\eta$ is a fixed threshold and $\nu$ controls sensitivity to contextual deviation.  The model learns a mapping $f: \overrightarrow{z_k} \rightarrow p_k$, where $p_k$ represents the predicted contextual behaviour of the target change. From the trained model, we extract attribute importance weights $\{w_1, w_2, \ldots, w_N\}$, where $w_j$ reflects the contribution of attribute $A_j$ to explain the changes in $A_t$. We extract the attributes with the highest weights, $\mathcal{R}_t$, representing attributes that most strongly influence changes in $A_t$.
% \fei{Give 1-2 sentences summary of how you identify and use these attributes in $\mathcal{R}_t$ in your search, so the reader has an idea instead of just saying go to the long paper.} \ns{To construct $\mathcal{R}_t$ we sort attributes by their learned weights and identify the largest consecutive drop in weight. All attributes preceding this largest gap are retained as $\mathcal{R}_t$.} \textcolor{green}{Further details can be found in the extended version~\cite{datasite}.}

To identify $\mathcal{R}_t$ we sort the weights for $R \setminus \{A_t\}$, in descending order $(w_{(1)} \ge w_{(2)} \ge \cdots \ge w_{(N-1)})$ and compute the differences as $\Delta_i = w_{(i)} - w_{(i+1)}$, $1 \le i < N-1$. The index of the largest drop is identified as $c^* = \arg \max_i \Delta_i$. This largest gap represents the point at which contextual relevance decreases most sharply. We therefore retain only the attributes preceding this gap as the relevant attribute set $\mathcal{R}_t = \{A_{(1)}, A_{(2)},\ldots, A_{(c^*)}\}$.

\noindent \textbf{Context factor.}
Using the learned weights, we define the context factor for tuple pair $(r_k, r_{k+1})$ as $\beta_k[A_t] = z_k[A_t] - \sum_{A_j \in \mathcal{R}_t} w_j \cdot z_k[A_j]$, measuring the deviation of the target attribute from its expected changes. We then scale the changes as: $\delta'_{(k,k+1)}[A_t] = \delta_{(k,k+1)}[A_t] \cdot \beta_k[A_t]$. We compute a sequence of context-aware change vectors and pass this into the rule mining stage.

% These adjusted changes are used for Change Rule discovery.
% \vspace{-0.1cm} 

% Thus, when mining change rules with $X = A_t$, \crminer evaluates candidate consequents only for $Y \in \mathcal{R}_t$, while for all other ordering attributes $X \neq A_t$, candidate consequents are drawn from the full set $R \setminus \{X\}$.

% Using the learned weights, we define the context factor $\beta_{(i,i+1)}[A']$ for a tuple pair $(r_i, r_{i+1})$ as the difference between the normalized change in the target and the weighted contributions of the non-target attributes:
% \[
% \beta = z_{(i,i+1)}[A'] - \sum_{A_k \in R\setminus A'} z_{(i,i+1)}[A_k] \cdot w_k
% \]
% This formulation ensures that $\beta_{(i,i+1)}[A']$ reflects how much the target’s change deviates from what is expected given the concurrent changes in the most relevant attributes. Finally, the observed change in the target attribute is augmented by $\beta_{(i,i+1)}[A']$, producing a context-aware change:
% \[
% \delta_{(i,i+1)}[ A'_{aug}] = \delta_{(i,i+1)}[A'] \cdot \beta_{(i,i+1)}[A']
% \]

% The output of this subsection is a sequence of consecutive changes for each tuple pair. The selected target attributes are context-adjusted while other attributes retain their raw changes. These change values serve as the input for dependency discovery. The next subsection formalizes how relationships between such changes are modeled and mined as Change Rules.

% \vspace{-0.7cm}

\section{\crminer}
    %This subsection defines Change Rules (CRs) and metrics used for validation. Then we define the mining algorithm, \crminer. The methodology consists of constructing differential functions from observed changes, encoding them using bitsets, and identifying valid change rules.
% \vspace{-0.5cm}

We introduce our CR mining algorithm, \crminer, by first defining change rules and then how we build diff-sets to identify candidate intervals encapsulating attribute changes.  We then use these intervals and incrementally enlarge them to identify non-redundant, minimal change rules. 

% \vspace{-0.3cm}

%\noindent \textbf{Change Rules.} Real-world datasets often exhibit consistent relationships between how one attribute changes and how another responds. To capture such relationships, we define Change Rules over consecutive tuple pairs.

\begin{definition} (Change Rules)
    Let $I$ be an instance ordered by a singular attribute $X \in R$, producing $\pi_X(I) = \langle r_1, \ldots, r_n \rangle$. A change rule $\phi$ \eat{captures a dependency between the changes in $X$ and a consequent attribute $Y \in R \setminus \{X\}$. It} is defined as $\phi: X_{g_{X}} \rightarrow_{g} Y$, where $g_X = [l_X, u_X]$ and $g = [l, u]$ are intervals that capture the minimum and maximum range of changes in $X$ and $Y$, respectively.  That is, we say $\phi$ holds for a consecutive pair (denoted as $(r_k, r_{k+1}) \models \phi$), if $\delta_{(k,k+1)}[X] \in g_X  \rightarrow \delta_{(k,k+1)}[Y] \in g$.
\end{definition}

\subsection{Building Diff-sets}

% \vspace{-0.4cm}

\begin{figure}[t]
    \centering
    \includegraphics[width=0.5\linewidth]{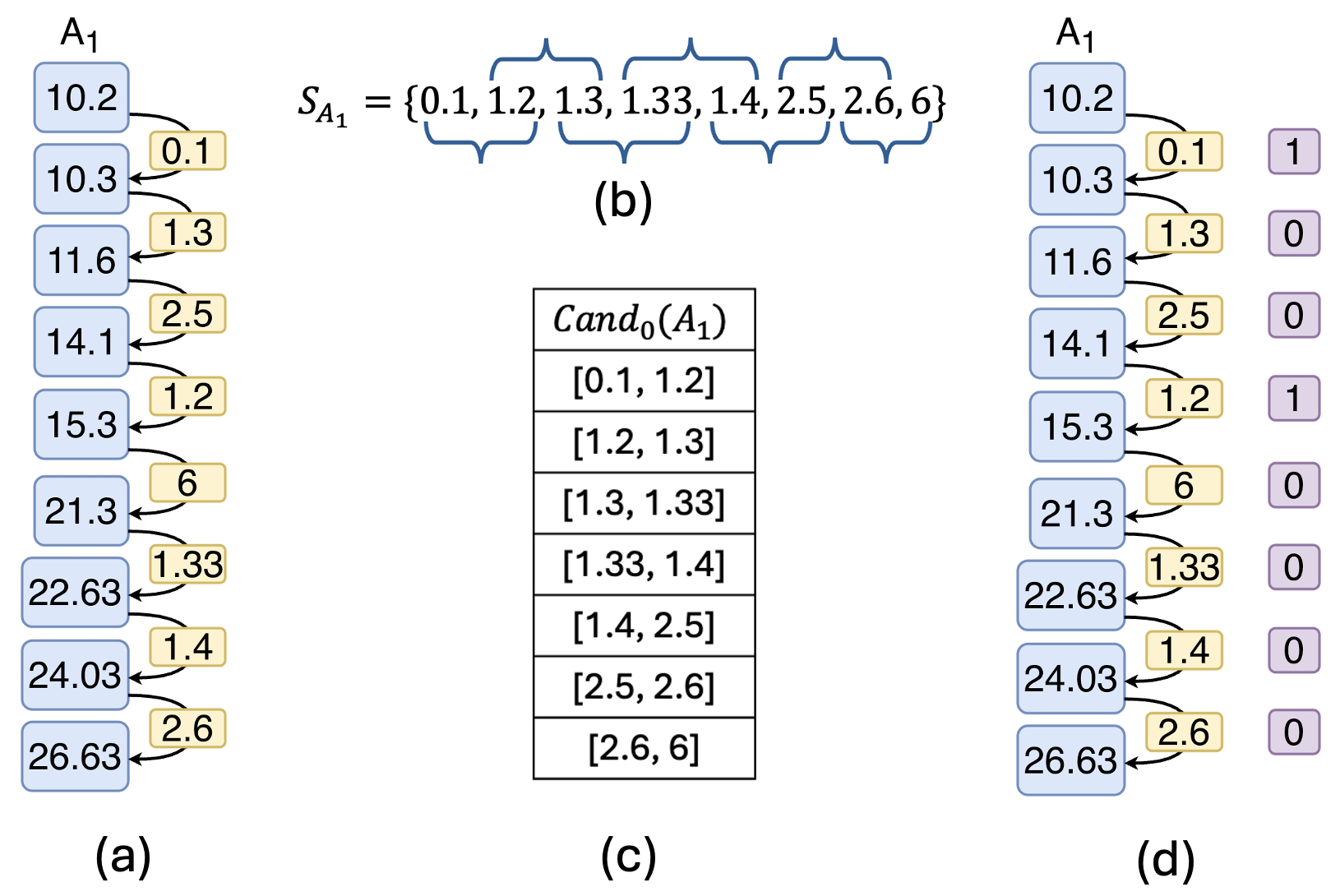}
    \caption{Generating candidate intervals and their bitsets}
    \label{fig:algoex}
    % \vspace{-0.4cm}
\end{figure}

A differential function on attribute $A_j$ is defined as $\lambda(A_j) = (\delta_{(k,k+1)} [A_j] \in [l,u])$\eat{, which is satisfied by a tuple pair $(r_k, r_{k+1})$ if the change in $A_j$ is within the range $[l,u]$}.  To construct a set of differential functions $\Psi$, we must do so incrementally.  For an attribute $A_j \in R$, let the sequence of changes between all tuple pairs of $\pi_X(I)$ be $\Delta A_j = \{\delta_{(1,2)}[A_j], \ldots, \delta_{(m-1,m)}[A_j]\}$, where each element in the set corresponds to a tuple pair $(r_k, r_{k+1})$. Let $S_{A_j} = \{s_1, s_2, \ldots\}$ be the sorted set of \textbf{distinct} change values in $\Delta A_j$. For example, from Figure~\ref{fig:algoex}(a), $S_{A_1} = \{0.1, 1.2, 1.3, 1.33, 1.4, 2.5, 2.6, 6\}$.

% \begin{figure}[htbp]
%     \centering
%     % First Minipage
%     \begin{minipage}{0.30\textwidth}
%         \centering
%         \includegraphics[height = 4cm]{Images/algoex2.png}
%         \caption{Consecutive changes in $A_1$}
%         \label{fig:algoex2}
%     \end{minipage}
%     \hfill
%     \begin{minipage}{0.30\textwidth}
%         \centering
%         \includegraphics[height = 3cm]{Images/algoex6.png}
%         \caption{Initial candidate intervals}
%         \label{fig:algoex6}
%     \end{minipage}
%     \hfill
%     \begin{minipage}{0.30\textwidth}
%         \centering
%         \includegraphics[height = 4cm]{Images/algoex3.png}
%         \caption{Bitset for the candidate interval $g_1 = [0.1,1.2]$}
%         \label{fig:algoex3}
%     \end{minipage}
% \end{figure}
% \vspace{-0.1cm}

We compute candidate intervals by starting with adjacent values in $S_{A_j}$, as shown in Figure~\ref{fig:algoex}(b).  The first set of candidate intervals for $A_1$ is shown in Figure~\ref{fig:algoex}(c). We proceed to increase the size of the intervals by incrementally expanding the upper bound.  Let the set of candidate intervals proceed in a levelwise manner starting at level 0, defined as $Cand_{0}(A_j) = \{[s_p, s_{p+1}] \;|\; s_p \in S_{A_j},  0 \le p \le |S_{A_j}|-1\}$.  For each interval $g_i \in Cand_0(A_j)$, we iterate through $\Delta A_j$ to form a bitset $\mathcal{B}(g_i)$ of length $|\Delta A_j|$, where:
\[
\mathcal{B}(g_i)[k] =
\begin{cases}
    1 & \text{if } \Delta A_j[k] \in g_i \\
    0 & \text{otherwise}
\end{cases}
\]
The $k$-th bit in the bitset represents \eat{whether the change in $A_j$ between the tuple pairs $(r_k, r_{k+1})$ falls within the range represented by $g_i$ (or }if $\delta_{(k,k+1)}[A_j] \in g_i$ or not, as shown in Figure~\ref{fig:algoex}(d). 
%At the end of this, we have a bitset $\mathcal{B}(g_i)$ for all $g_i \in Cand_0(A_j)$.
We iteratively construct larger candidate intervals by combining adjacent intervals from the previous level. For level $\ell > 0$, candidate intervals are generated as: $Cand_\ell(A_j) = \{[s_p, s_{p+\ell+1}] \;|\; 0 \le p \le |S_{A_j}| - \ell - 2\}$. Instead of recomputing bitsets, we exploit the composability of adjacent intervals. For two consecutive intervals $g_a = [s_p, s_q]$ and $g_b = [s_q, s_r]$, the bitset of their merged intervals can be computed as $\mathcal{B}(g_a \cup g_b) = \mathcal{B}(g_a) \;\lor\; \mathcal{B}(g_b)$. For example, $\mathcal{B}([0.1,1.3]) = \mathcal{B}([0.1,1.2])\;\lor\;\mathcal{B}([1.2,1.3] = 10010000\;\lor\;01010000 = 11010000$.  For each candidate interval $g_i$, we compute the length of the longest consecutive subsequence of 1's in its bitset $\mathcal{B}(g_i)$. Let this length be denoted as $max\_seg(g_i)$. 

% \vspace{-0.3cm}

\begin{definition} (Segment Coverage)
    The segment coverage of a candidate interval $g_i$ is defined as the fraction of consecutive tuple pairs contained in its longest contiguous segment, i.e., $coverage(g_i) = \frac{max\_seg(g_i)}{m - 1}$.
\end{definition}

% \vspace{-0.3cm}

\eat{The support of $g_i$ is defined as $supp(g_i) = \frac{max\_seg(g_i)}{m - 1}$.} To compute $max\_seg(g_i)$ efficiently, we scan $\mathcal{B}(g_i)$ from left to right and only initiate counting at positions where the bit is 1. For each such position $t$, we extend forward to measure the length of the consecutive run of 1's starting at $t$. Let $\tau = \lceil \theta_c \cdot (m - 1) \rceil$ be the minimum required segment length to satisfy the \eat{support} segment coverage threshold $\theta_c$. During the scan, if the current starting position is $t$, then even in the best case (i.e., all remaining bits are 1), the maximum possible segment length is $(m - 1 - t)$. Therefore, if $(m - 1 - t) < \tau$, we can terminate early, as no valid segment can be found beyond this point. If no segment of length at least $\tau$ is found, the candidate interval $g_i$ is pruned.  If a candidate interval $g_i$ satisfies $coverage(g_i) \ge \theta_c$, it is added to the set of differential functions for attribute $A_j$, denoted as $\psi_j$.

\subsection{Mining Change Rules} 

% \vspace{-0.3cm}

The set of differential functions across all attributes is denoted as $\Psi = \{\psi_{A_1}, \psi_{A_2},\\ \ldots, \psi_{A_N}\}$, where $\psi_X \in \Psi$ contains the differential functions for the antecedent attribute $X$. Given an interval $g_X \in \psi_X$, \crminer identifies a differential function $g \in \psi_Y$ for $Y \in R \setminus \{X\}$.  When $X$ is equivalent to the attribute $A_t$, the candidate consequent attribute is restricted to $\mathcal{R}_t$, the context-relevant attributes found during the previous (context-aware change) stage. A candidate change rule is discovered based on the co-occurrence of the longest valid segments of the differential functions of $X$ and $Y$, such that it meets the minimum support threshold $\theta$.

% For a general ordering attribute $X$, all attributes in $R \setminus \{X\}$ are considered as candidate consequents. However, when $X$ is the target attribute $A_t$, the search is restricted to a context-aware subset of relevant attributes selected using the learned contextual relevance weights. That is, only attributes with the strongest influence on changes in $A_t$ are retained as candidate RHS attributes. 

% \vspace{-0.5cm}

\begin{definition} (Change Rule Support)
    The support of a change rule $\phi: X\;\;_{g_X} \rightarrow_g Y$ defined over an instance $I$ measures the proportion of tuple pairs that satisfy $\phi$. Let $\mathcal{S} = \{ (r_i, r_{i+1}) \in I \mid (r_i,r_{i+1}) \models \phi \}$, such that $|\mathcal{S}| \geq |\mathcal{S'}|, \forall \mathcal{S'} \subseteq \mathcal{S}$. We define $support(\phi)=\frac{|\mathcal{S}|}{|I|-1}$.
\end{definition}

A CR $\phi$ is said to hold with threshold $\theta$ if $support(\phi) \ge \theta$.

\noindent \textbf{Algorithms.}  We now describe the details of \crminer, which is divided into two main algorithms: (i) BuildDiff, and (ii) Discover Change Rules.

\noindent \underline{BuildDiff (Alg.~\ref{algo:df})}: Constructs differential functions $\Psi$ and bitsets $\mathcal{B}$ via level-wise candidate generation. For each attribute $A_j \in R$, consecutive changes $\Delta A_j$ are sorted into distinct values $S_{A_j}$ (Line 4). Level-0 candidates $Cand_0(A_j)$ are initialized from adjacent values in $S_{A_j}$, and bitsets are built to mark satisfying tuple pairs (Line 6). Iterating level-wise (Line 8), each candidate interval $g \in Cand_\ell$ is evaluated by measuring its bitset's longest consecutive run of 1's to determine segment coverage (Lines 9--18). Intervals satisfying threshold $\theta_c$ are appended to $\psi_{A_j}$ (Lines 19--20). Remaining positions unable to satisfy the minimum segment length $\tau$ trigger early termination via upward pruning (Line 20). Next-level candidates are generated by merging adjacent intervals, with bitsets computed efficiently via bitwise OR (Lines 21--27). The loop terminates when no candidates remain, returning $\Psi$ and $\mathcal{B}$.

\begin{algorithm}[H]
    \LinesNumbered
    \SetAlgoLined
    \caption{BuildDiff}
    \label{algo:df}
    \KwIn{Ordered tuple sequence $\pi_X(I)$, segment coverage threshold $\theta_c$}
    \KwOut{Differential functions $\Psi$ and bitsets $\mathcal{B}$}

    $\Psi \leftarrow \emptyset$, $\mathcal{B} \leftarrow \emptyset$

    $\tau \leftarrow \lceil \theta_c \cdot (m-1) \rceil$

    \ForEach{$A_j \in R$}
    {
        $S_{A_j} \leftarrow$ sorted distinct values in $\Delta A_j, \;\psi_{A_j} \leftarrow \emptyset$
        
        $Cand_0 \leftarrow \{[s_p, s_{p+1}] \mid 0 \le p \le |S_{A_j}| - 2\}$

       compute $\mathcal{B}(g)$ for each $g \in Cand_0$

        $\ell \leftarrow 0$

        \While{$Cand_\ell \neq \emptyset$}
        {
            $Cand_{\ell+1} \leftarrow \emptyset$

            \ForEach{$g \in Cand_\ell$}
            {
                $current\_run \leftarrow 0$, $max\_seg(g) \leftarrow 0$
                $expandable(g) \leftarrow \text{true}$

                \For{$t = 1$ to $m-1$}
                {
                    break if $(m - t) < \tau$

                    \If{$\mathcal{B}(g)[t] == 1$}
                    {
                        $current\_run \leftarrow current\_run + 1$

                        $max\_seg(g) \leftarrow \max(max\_seg(g), current\_run)$
                    }
                    \Else
                    {
                        $current\_run \leftarrow 0$
                    }
                }

                \If{$coverage(g) \leftarrow \frac{max\_seg(g)}{m-1} \ge \theta_c$}
                {
                    $\psi_{A_j} \leftarrow \psi_{A_j} \cup \{g\},\; \mathcal{B}[g] \leftarrow \mathcal{B}(g), \;expandable(g) \leftarrow \text{false}$
                }
            }

            \For{$i = 0$ to $|Cand_\ell| - 2$}
            {
                $g_a \leftarrow Cand_\ell[i], \;g_b \leftarrow Cand_\ell[i+1]$

                \If{$expandable(g_a)$ \textbf{and} $expandable(g_b)$}
                {
                    $g_{new} \leftarrow [l(g_a), u(g_b)]$

                    $\mathcal{B}(g_{new}) \leftarrow \mathcal{B}(g_a) \lor \mathcal{B}(g_b)$

                    add $g_{new}$ to $Cand_{\ell+1}$
                }
            }

            $\ell \leftarrow \ell + 1$
        }

        $\Psi \leftarrow \Psi \cup \psi_{A_j}$
    }

    \Return $\Psi$, $\mathcal{B}$
\end{algorithm}
% \vspace{-1cm}

% % \vspace{-0.3cm}

% \vspace{-0.7cm}

% \vspace{-1cm}
\noindent \underline{Discover Change Rules (Alg.~\ref{algo:cr})}: Extracts valid change rules $\Sigma$ using $\Psi$ and $\mathcal{B}$ as inputs. It first extracts antecedent differential functions $\psi_X$ for attribute $X$ (Line 2). If $X = A_t$, the consequent candidates are restricted to $\mathcal{R}_t$; otherwise, all attributes in $R \setminus \{X\}$ are considered (Lines 3--6). For each antecedent interval $g_X \in \psi_X$, consequent attribute $Y \in \mathcal{C}$, and consequent interval $g \in \psi_Y$, the algorithm intersects $\mathcal{B}(g_X)$ and $\mathcal{B}(g)$ via bitwise AND to find the longest consecutive run of 1's (Lines 12--20). Rule support is calculated, and if $support(\phi) \ge \theta$ and it passes the minimality check, $\phi$ is added to $\Sigma$ (Lines 21--28). The algorithm iterates through all pairs and returns $\Sigma$.

\begin{algorithm}[H]
    \LinesNumbered
    \SetAlgoLined
    \caption{Discover Change Rules}
    \label{algo:cr}
    \KwIn{Differential functions $\Psi$, bitsets $\mathcal{B}$, ordering attribute $X$, thresholds $\theta_c$, $\theta$, support tolerance $\epsilon_s$, target attribute $A_t$, and set of corresponding relevance weights $\{w_j\}$ for each $A_j$}
    \KwOut{Change Rule set $\Sigma$}
    \setcounter{AlgoLine}{0}

    $\Sigma \leftarrow \emptyset$

    $\psi_X \leftarrow$ differential functions for attribute $X$
    
    \eIf{$X = A_t$}
    {
        % sort weights: $w_{(1)} \ge w_{(2)} \ge \cdots \ge w_{(n)}$, $n = |R| - 1$\
        
        % $c^* \leftarrow \arg\max_i \,(w_{(i)} - w_{(i+1)})$\
        
        $\mathcal{C} \leftarrow \mathcal{R}_t$
    }
    {
        $\mathcal{C} \leftarrow R \setminus \{X\}$
    }

    \ForEach{$g_X \in \psi_X$}
    {
        $b_X \leftarrow \mathcal{B}(g_X)$

       \ForEach{$Y \in \mathcal{C}$}
        {
            $\psi_Y \leftarrow$ differential functions for attribute $Y$
            
            \ForEach{$g \in \psi_Y$}
            {
                $b_Y \leftarrow \mathcal{B}(g)$
                
                $b_{AND} \leftarrow b_X \land b_Y$
                
                $current\_run \leftarrow 0$, $max\_seg \leftarrow 0$
                
                \For{$t = 1$ to $|I|-1$}
                {
                    \eIf{$b_{AND}[t] == 1$}
                    {
                        $current\_run \leftarrow current\_run + 1$
                        
                        $max\_seg \leftarrow \max(max\_seg,\, current\_run)$
                    }
                    {
                        $current\_run \leftarrow 0$
                    }
                }
                \If{$support(\phi) \leftarrow \dfrac{max\_seg}{|I|-1} \ge \theta$}
                {
                    $minimal \leftarrow \text{true}$
                    
                    \ForEach{$\phi' \in \Sigma$ over same $(X, Y)$}
                    {
                        \If{$g'_X \text{ is contained in } g_X$ \textbf{or} $g' \text{ is contained in } g$}
                        {
                            $minimal \leftarrow \text{false}$
                            
                            \textbf{break}
                        }
                    }
                    \If{$minimal$}
                    {
                        $\Sigma \leftarrow \Sigma \cup \{\phi: X\;\;_{g_X} \rightarrow_g Y\}$
                    }
                }
            }
        }
    }
    \Return $\Sigma$
\end{algorithm}

\section{Experiments}

\stitle{Setup Environment.}
We implement \crminer using Python 3.10.18 and conducted experiments in a local Conda environment, macOS with an M1 chip and 8 GB RAM. Default parameters are set to $\theta = 0.9$, $\theta_c = 0.95$, $\epsilon_s = 0.01$, and $\mathcal{W} = 8$. For comparative experiments, we evaluate \crminer against FastDD, an existing DD discovery baseline algorithm that also models bounded differences using diff-sets ~\cite{kuang2024efficient}. All DDs reported in our evaluation are found using FastDD. \crminer extends the FastDD framework to ordered sequential data, whereas DDs are defined over all tuple pairs in the relation.

% % \vspace{-0.5cm}

% We acknowledge that FastDD differs in its assumptions, particularly in not incorporating tuple ordering or context-aware scaling. Therefore, the comparison is intended to evaluate the efficiency of \crminer relative to the closest existing DD mining approach, rather than a strict one-to-one comparison of identical dependency semantics.

%The window size for each experiment is fixed at 4.
%Both environments provide a robust infrastructure to validate the performance and scalability of our Parallel-TGFDMiner.

\begin{figure}[t]
    \centering
    \begin{subfigure}[b]{0.24\linewidth}
        \centering
        \includegraphics[width=\linewidth]{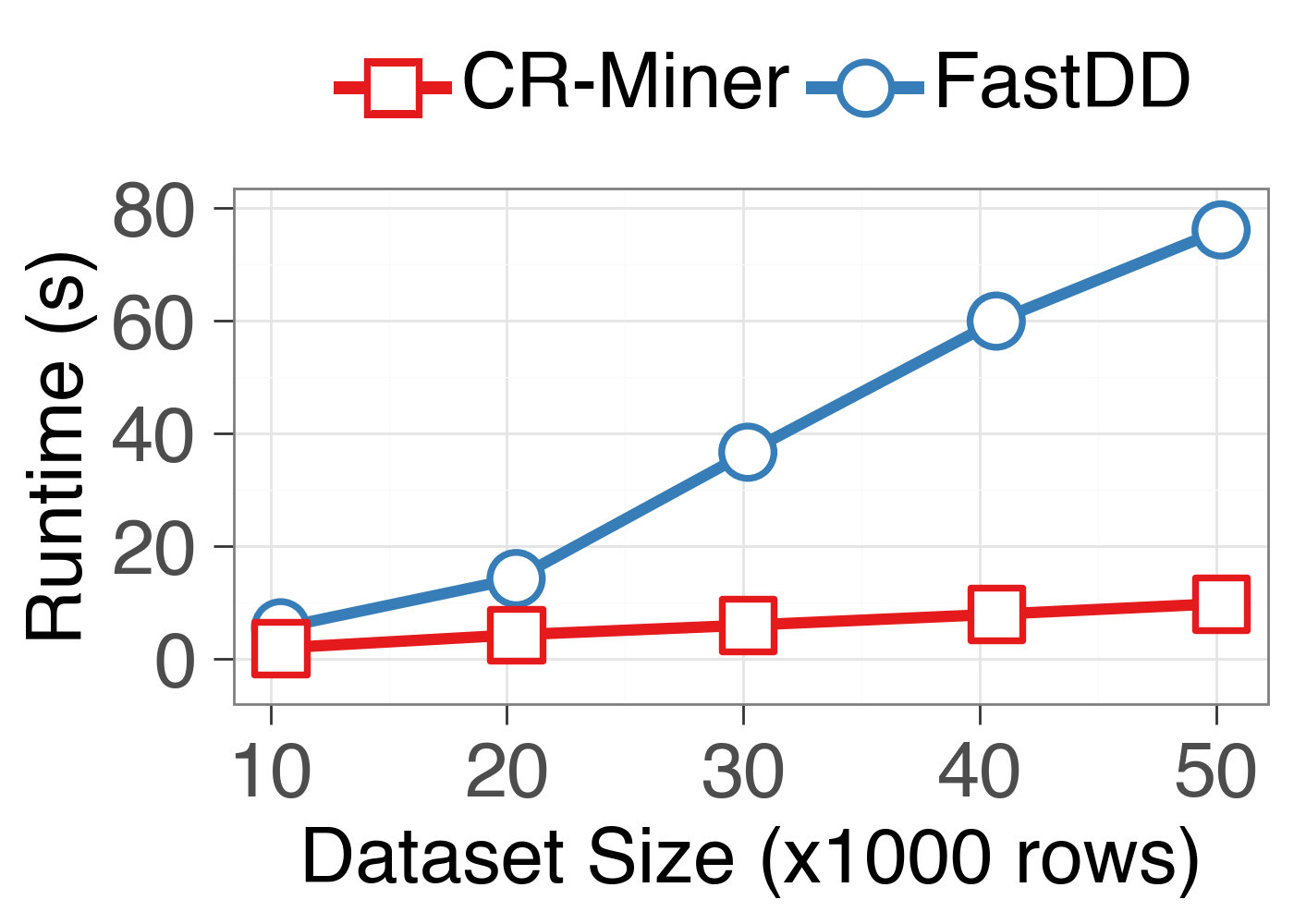}
        \caption{MIMIC-III}
        \label{fig:mimicexp1}
    \end{subfigure}
    \hfill
    \begin{subfigure}[b]{0.24\linewidth}
        \centering
        \includegraphics[width=\linewidth]{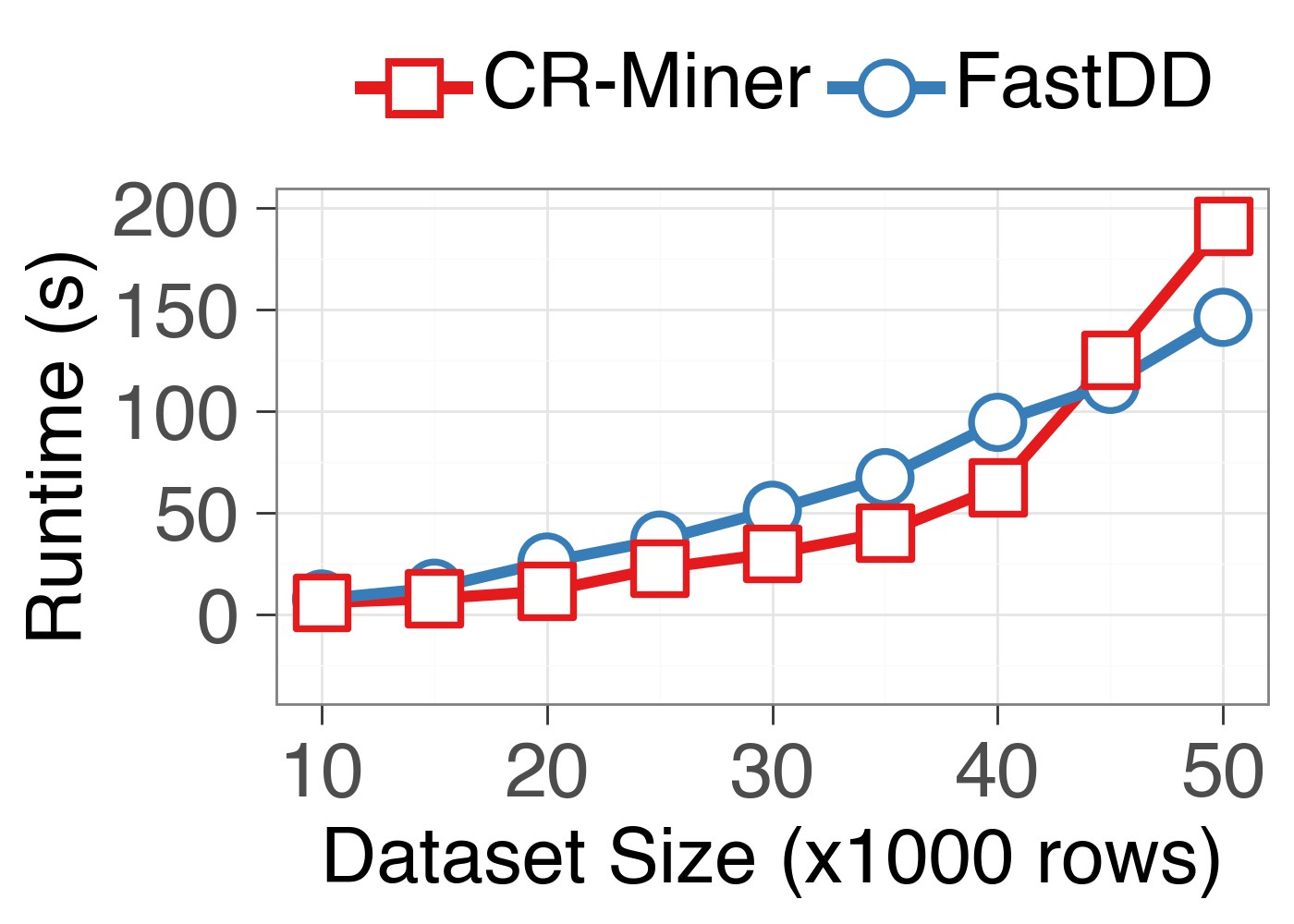}
        \caption{Employment}
        \label{fig:empexp1}
    \end{subfigure}
    \hfill
    \begin{subfigure}[b]{0.24\linewidth}
        \centering
        \includegraphics[width=\linewidth]{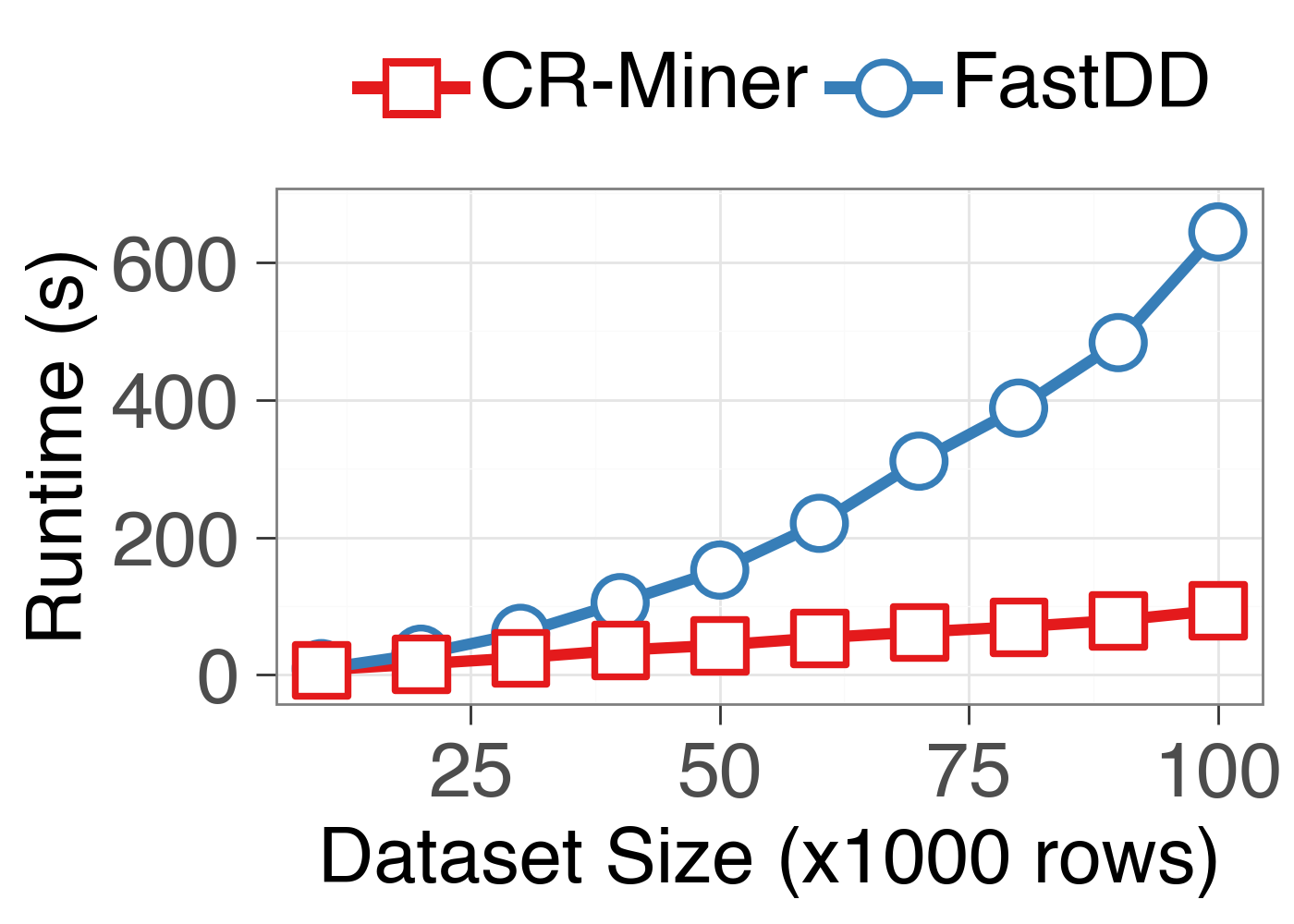}
        \caption{Weather}
        \label{fig:weatherexp1}
    \end{subfigure}
    \hfill
    \begin{subfigure}[b]{0.24\linewidth}
        \centering
        \includegraphics[width=\linewidth]{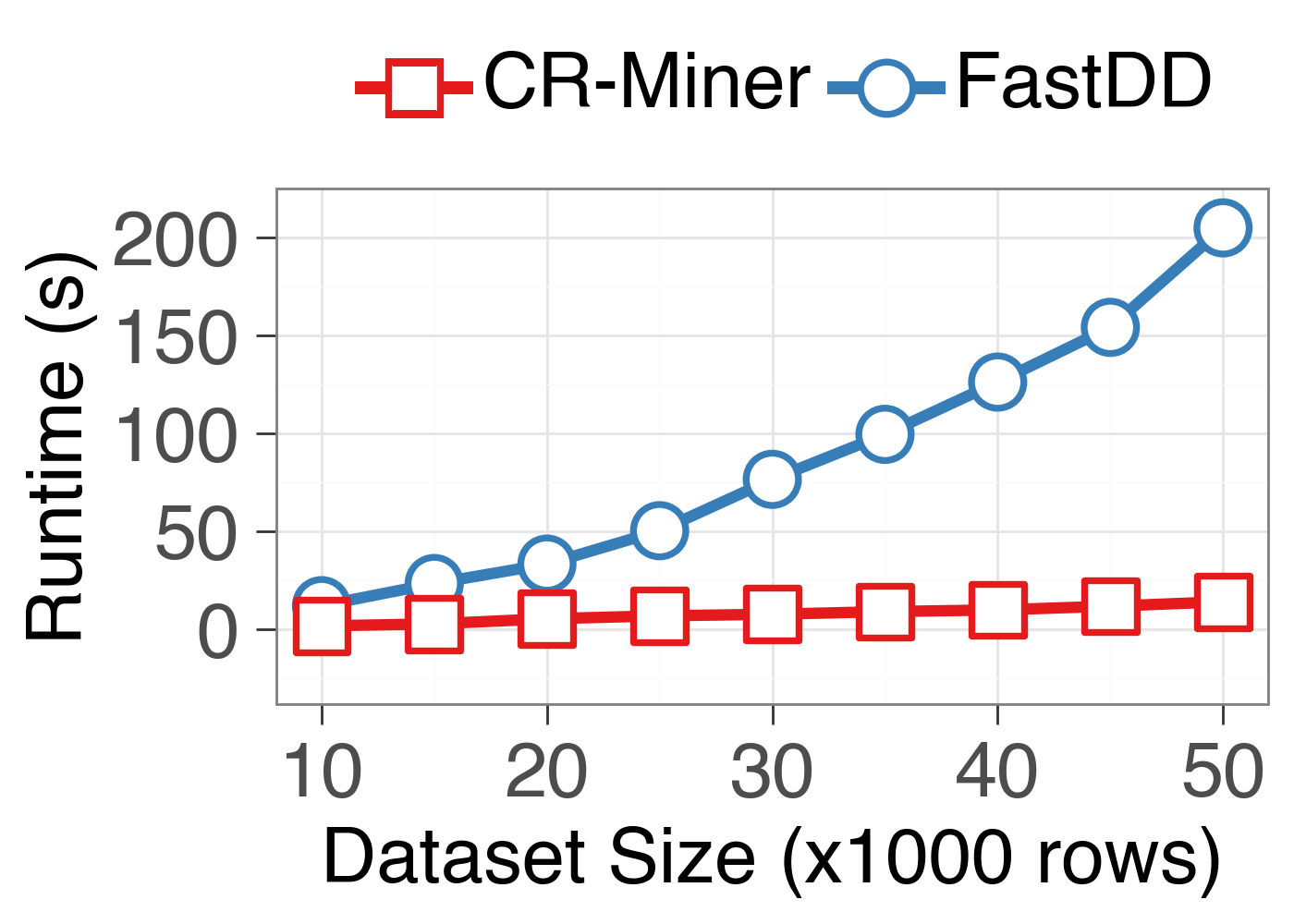}
        \caption{Power}
        \label{fig:powerexp1}
    \end{subfigure}
    \hfill
    \begin{subfigure}[b]{0.24\linewidth}
        \centering
        \includegraphics[width=\linewidth]{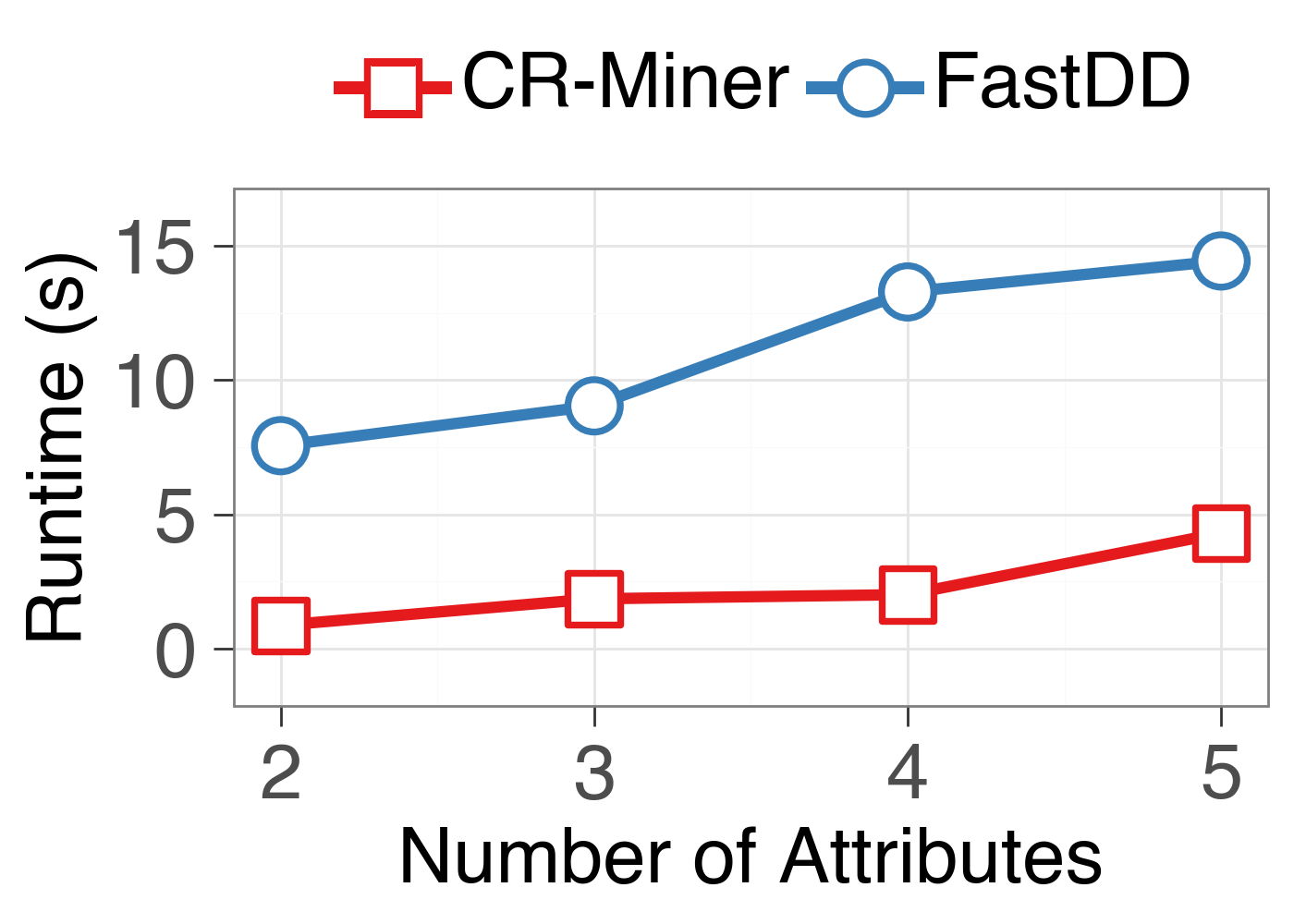}
        \caption{MIMIC-III}
        \label{fig:mimicexp3}
    \end{subfigure}
    \hfill
    \begin{subfigure}[b]{0.24\linewidth}
        \centering
        \includegraphics[width=\linewidth]{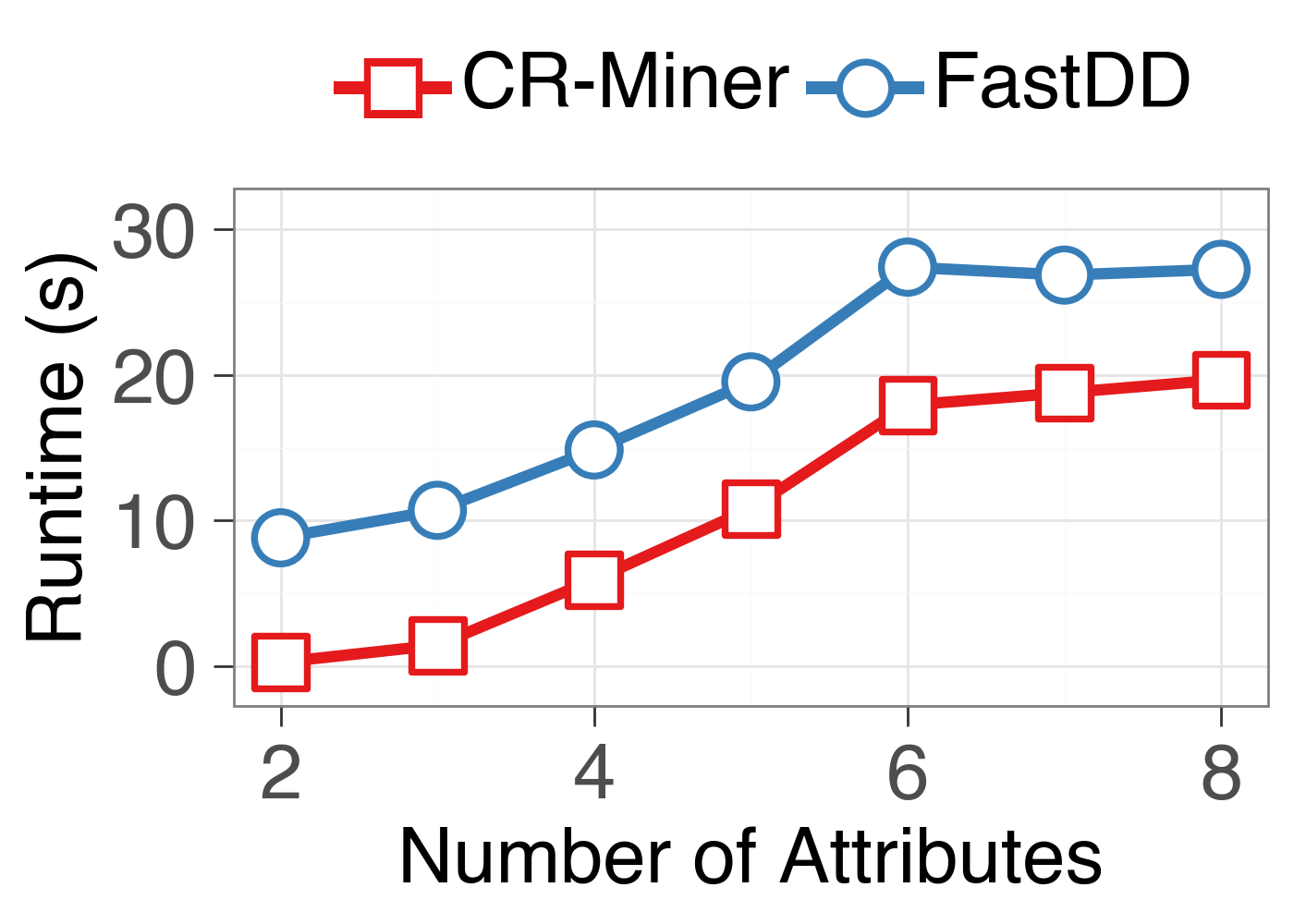}
        \caption{Employment}
        \label{fig:empexp3}
    \end{subfigure}
    \hfill
    \begin{subfigure}[b]{0.24\linewidth}
        \centering
        \includegraphics[width=\linewidth]{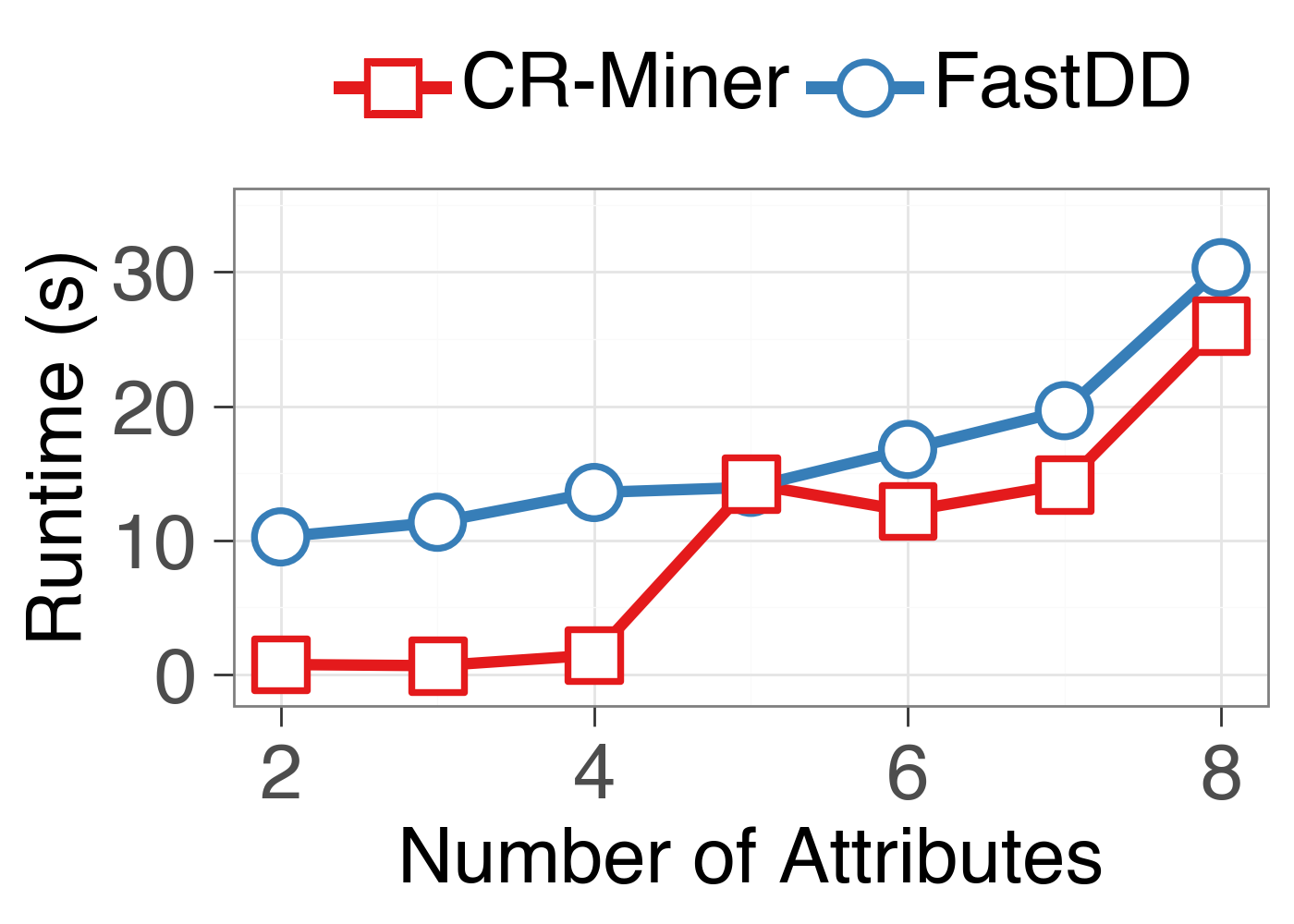}
        \caption{Weather}
        \label{fig:weatherexp3}
    \end{subfigure}
    \hfill
    \begin{subfigure}[b]{0.24\linewidth}
        \centering
        \includegraphics[width=\linewidth]{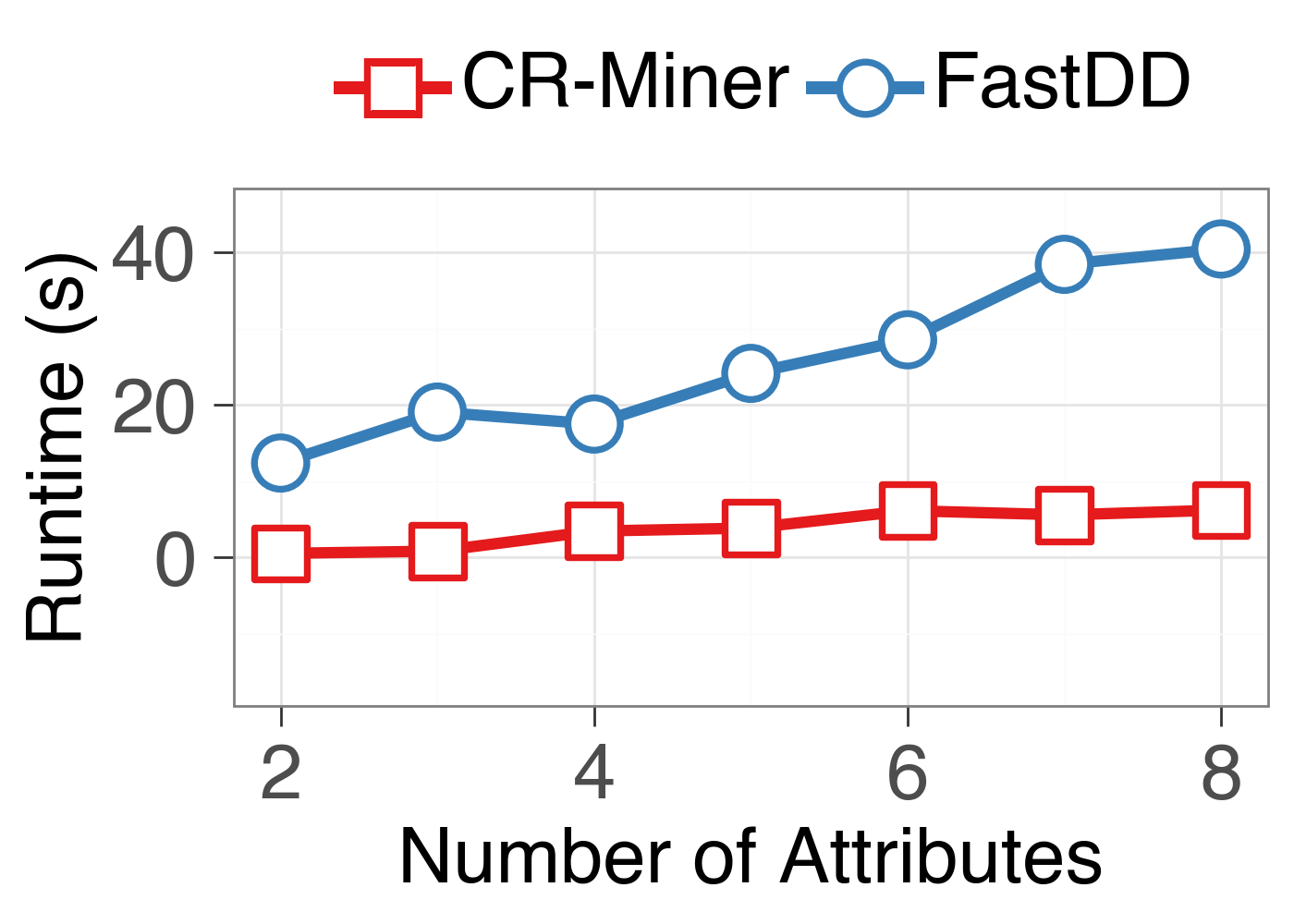}
        \caption{Power}
        \label{fig:powerexp3}
    \end{subfigure}
    \caption{Comparative runtime evaluation}
    \label{fig:overall_runtime}
\end{figure}

\stitle{Datasets.}
We use three real-world datasets covering healthcare, employment, and weather domains.  Our data and source code are publicly available~\cite{datasite}. \\
(1) \uline{MIMIC-III}\cite{johnson2016mimic}:  describes the healthcare information of patients admitted to the emergency department and ICU at a hospital in Boston, MA, from 2001-2012.  We focus on laboratory results covering 5 attributes and patient demographics over 50K records.  \\
(2) \uline{Employment}~\cite{uscensus}: describes industry-wide employment distributions across the US states, describing attributes such as firm size, total payroll, and total number of employees.  Our dataset contains 8 attributes and 50k records spanning employment statistics from 2012 to 2022. \\
(3) \uline{Weather}~\cite{weather}: this data is collected from the Environment and Climate Change Canada daily weather observations for the province of Ontario, spanning 2015 to 2025. The data contains provincial meteorological measurements such as temperature and precipitation.  Our data includes 8 attributes and 100k records. \\
(4) \uline{Power}~\cite{power}: this data describes the power consumption of three different distribution networks of Tetouan city, Morocco, in 2017. The data includes meteorological measurements and the power consumed in each zone every 10 minutes. It contains 8 attributes and more than 52k records.

\subsection{Experimental Results}
% \vspace{-0.3cm}

\noindent \textbf{Exp-1: Runtime vs. data size.} Figure~\ref{fig:overall_runtime} (a)-(d) shows the runtimes of \crminer and FastDD as the dataset size increases across the three datasets. \crminer scales more efficiently on MIMIC-III, Weather, and Power, exhibiting near-linear growth with significantly lower runtime compared to FastDD. In contrast, FastDD shows a steeper increase, indicating higher sensitivity to dataset size due to pairwise comparisons. On the Employment dataset, \crminer initially outperforms FastDD but exhibits a sharper increase at larger sizes, eventually exceeding FastDD at larger data sizes. This suggests that runtime is also influenced by attribute characteristics and interval generation complexity.

\begin{figure}[t]
    \centering
    \begin{subfigure}[b]{0.24\linewidth}
        \centering
        \includegraphics[width=\linewidth]{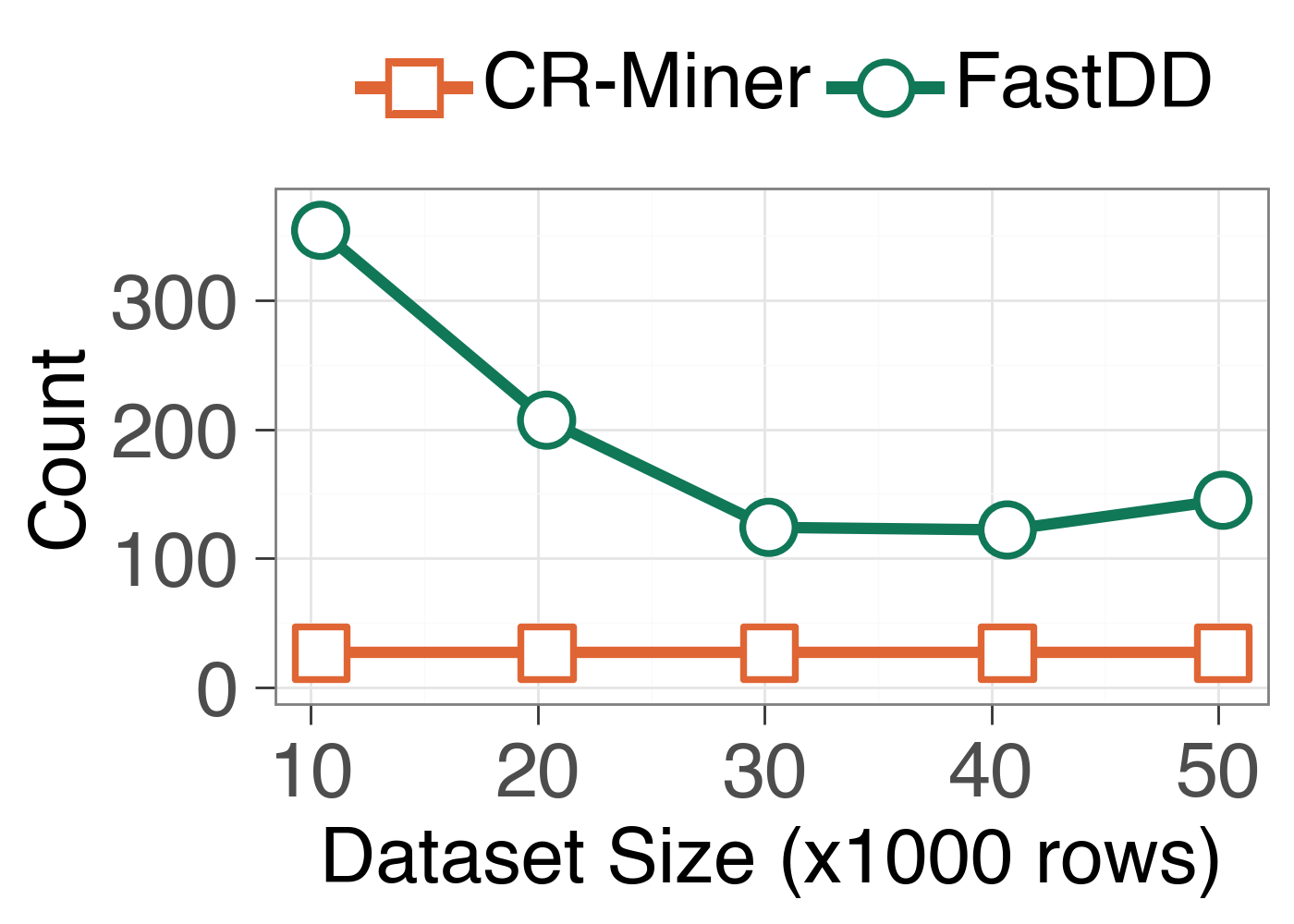}
        \caption{MIMIC-III}
        \label{fig:mimicexp2}
    \end{subfigure}
    \hfill
    \begin{subfigure}[b]{0.24\linewidth}
        \centering
        \includegraphics[width=\linewidth]{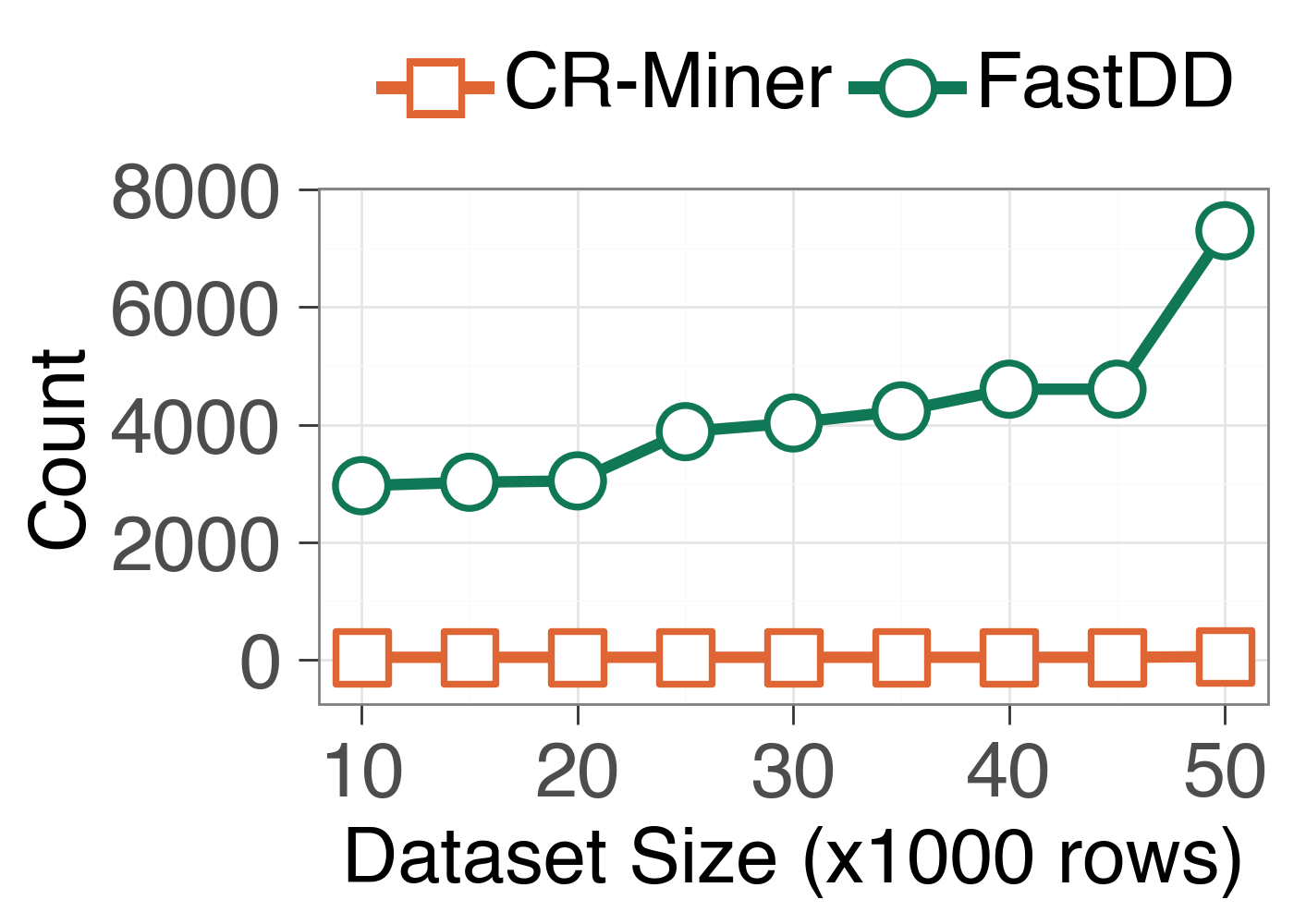}
        \caption{Employment}
        \label{fig:empexp2}
    \end{subfigure}
    \hfill
    \begin{subfigure}[b]{0.24\linewidth}
        \centering
        \includegraphics[width=\linewidth]{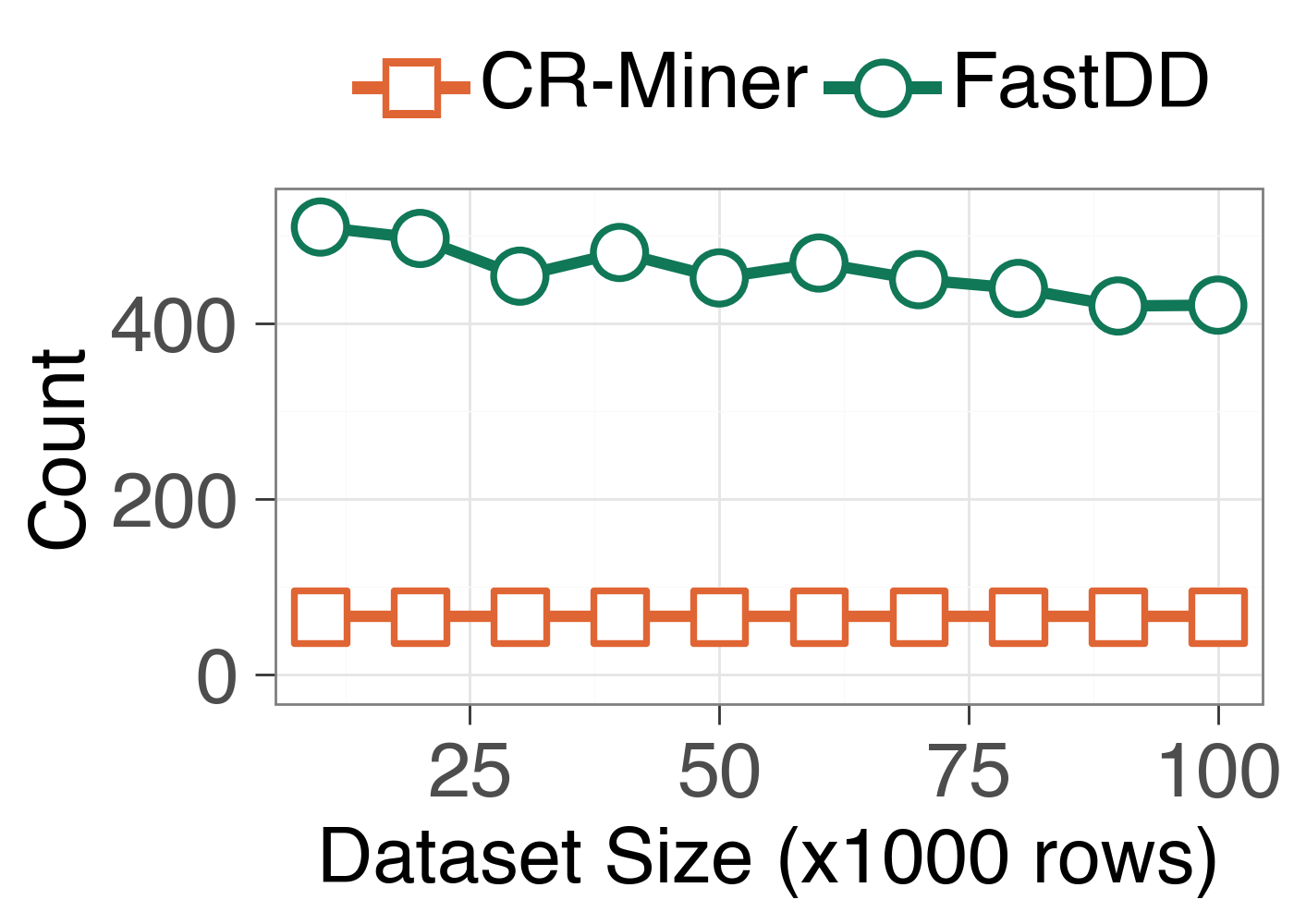}
        \caption{Weather}
        \label{fig:weatherexp2}
    \end{subfigure}
    \hfill
    \begin{subfigure}[b]{0.24\linewidth}
        \centering
        \includegraphics[width=\linewidth]{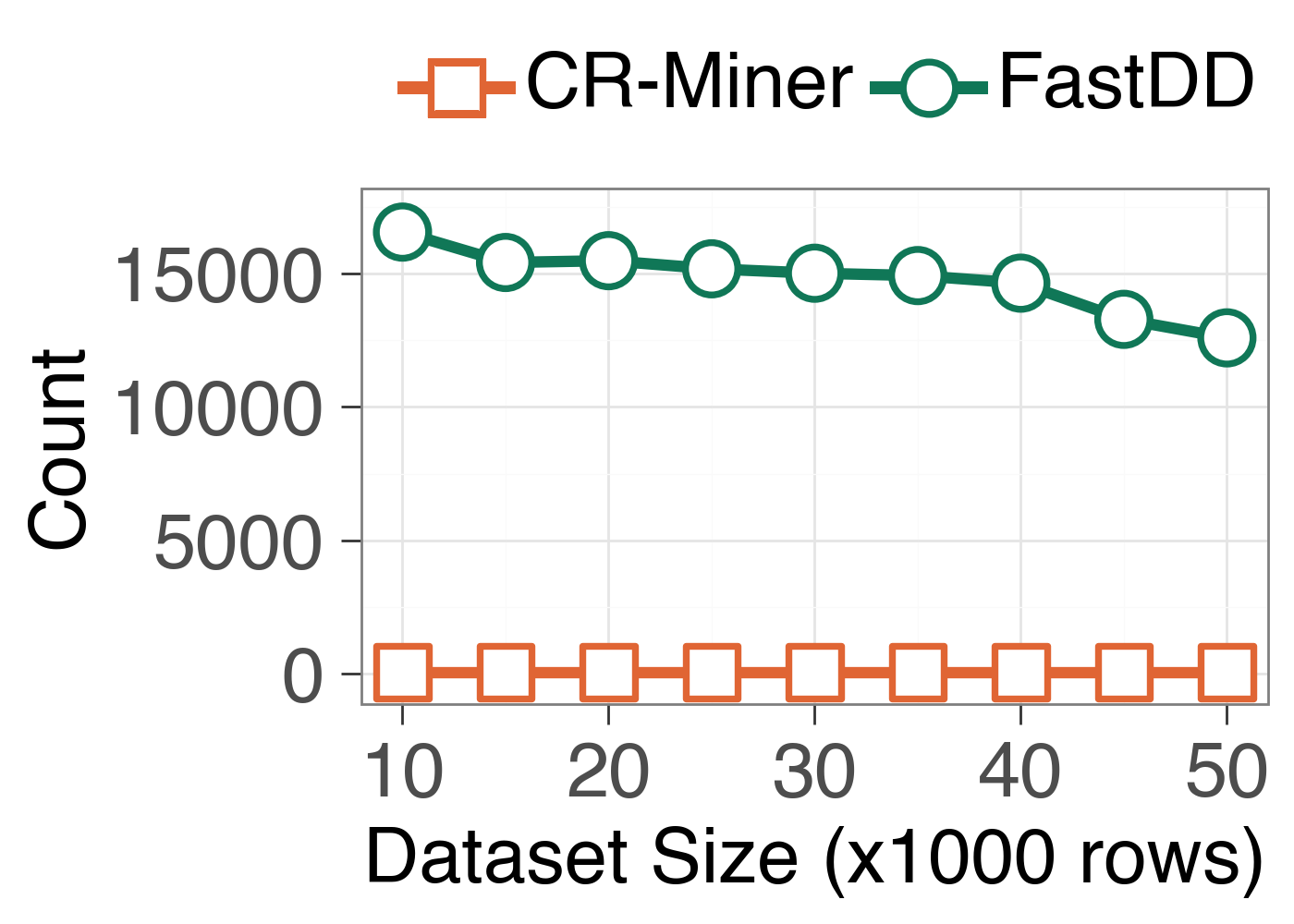}
        \caption{Power}
        \label{fig:powerexp2}
    \end{subfigure}
    \hfill
    \begin{subfigure}[b]{0.24\linewidth}
        \centering
        \includegraphics[width=\linewidth]{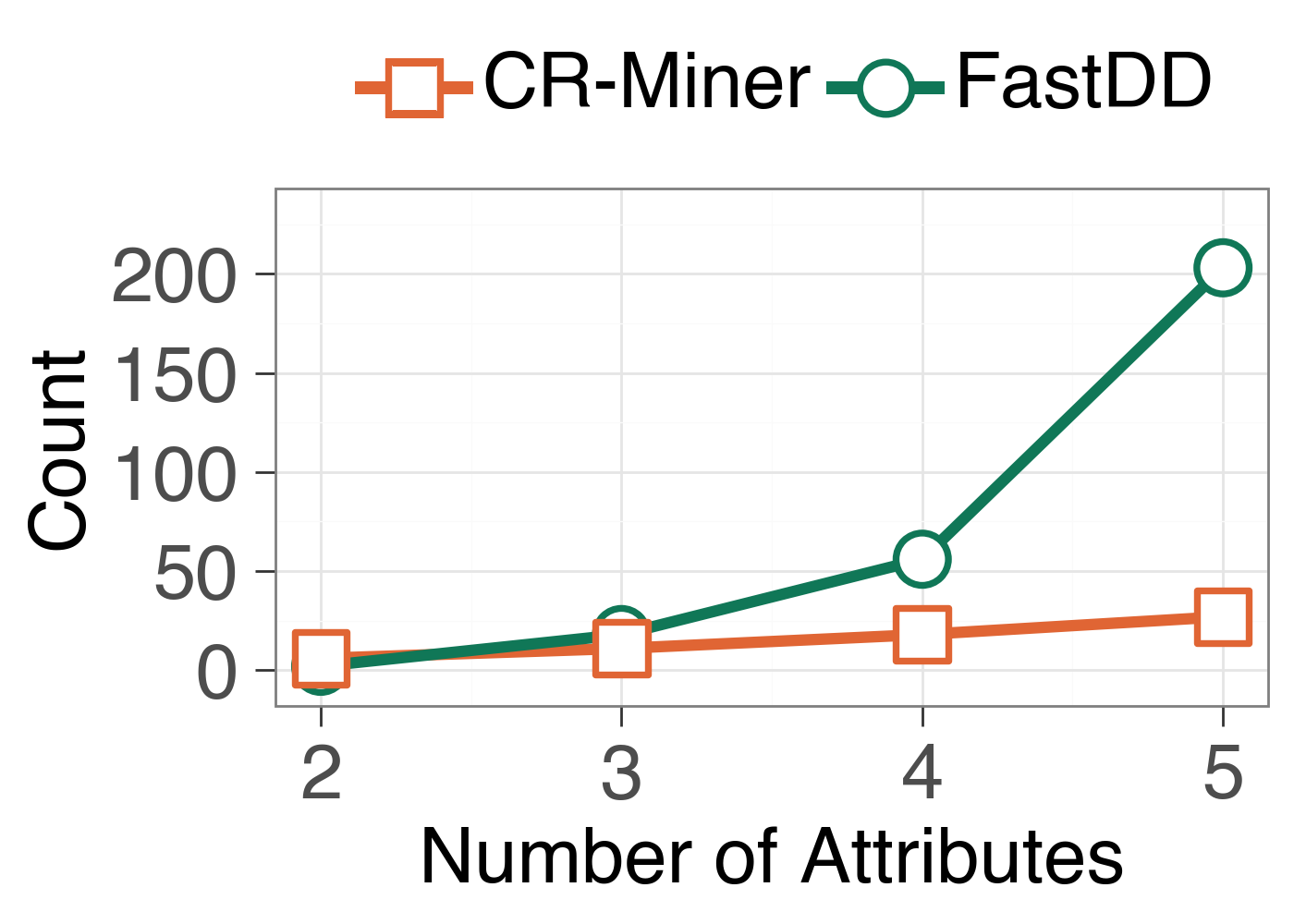}
        \caption{MIMIC-III}
        \label{fig:mimicexp4}
    \end{subfigure}
    \hfill
    \begin{subfigure}[b]{0.24\linewidth}
        \centering
        \includegraphics[width=\linewidth]{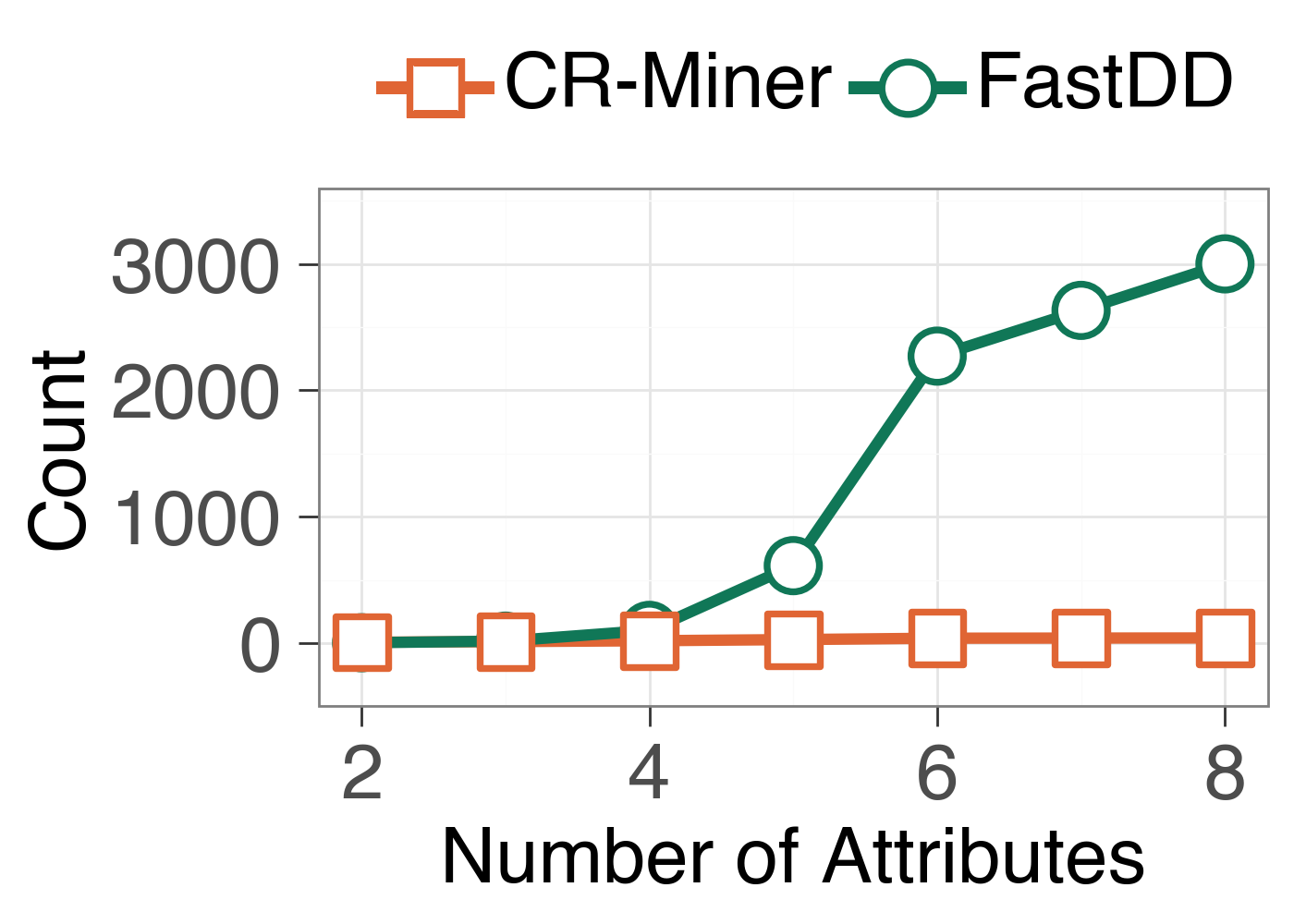}
        \caption{Employment}
        \label{fig:empexp4}
    \end{subfigure}
    \hfill
    \begin{subfigure}[b]{0.24\linewidth}
        \centering
        \includegraphics[width=\linewidth]{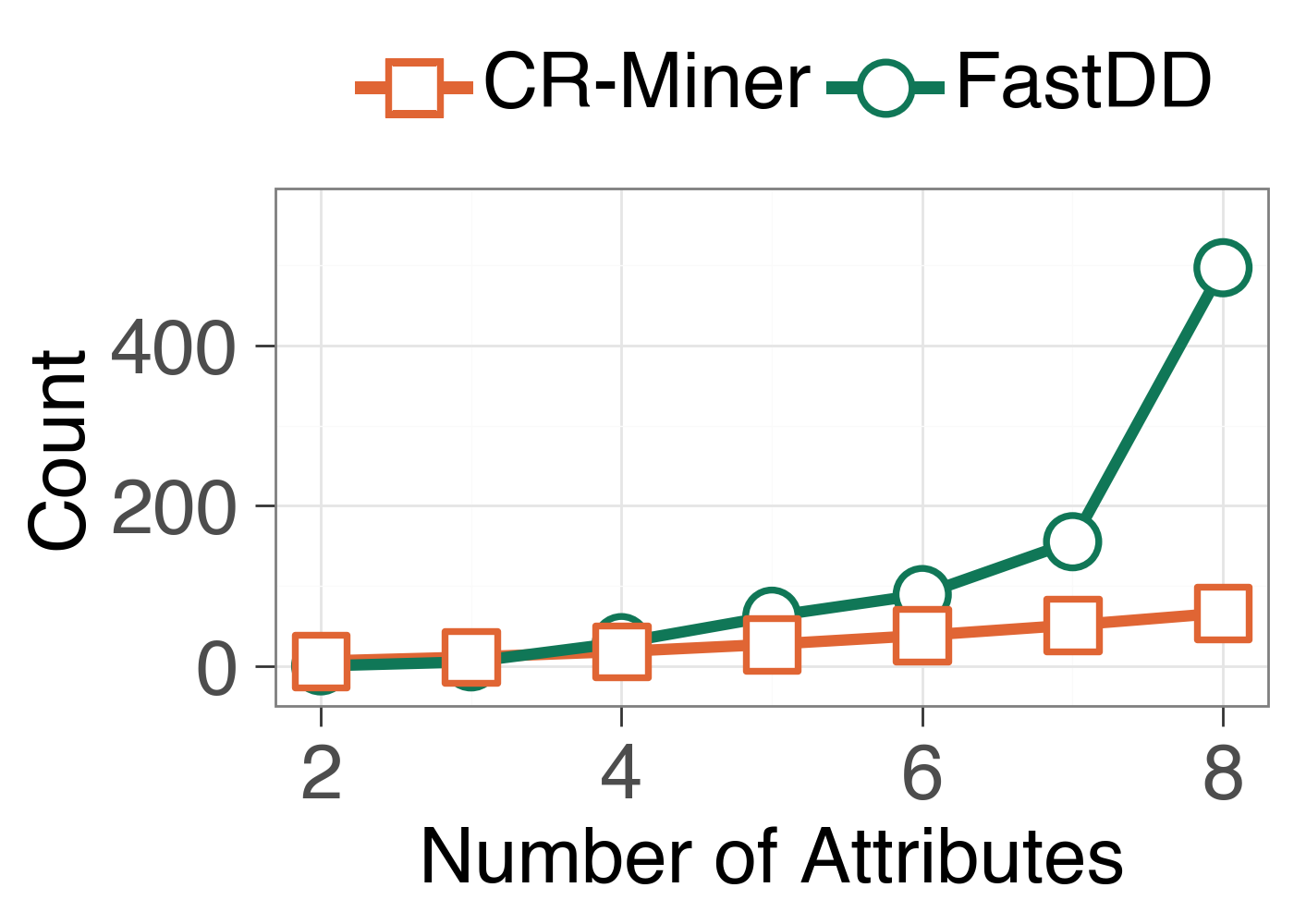}
        \caption{Weather}
        \label{fig:weatherexp4}
    \end{subfigure}
    \hfill
    \begin{subfigure}[b]{0.24\linewidth}
        \centering
        \includegraphics[width=\linewidth]{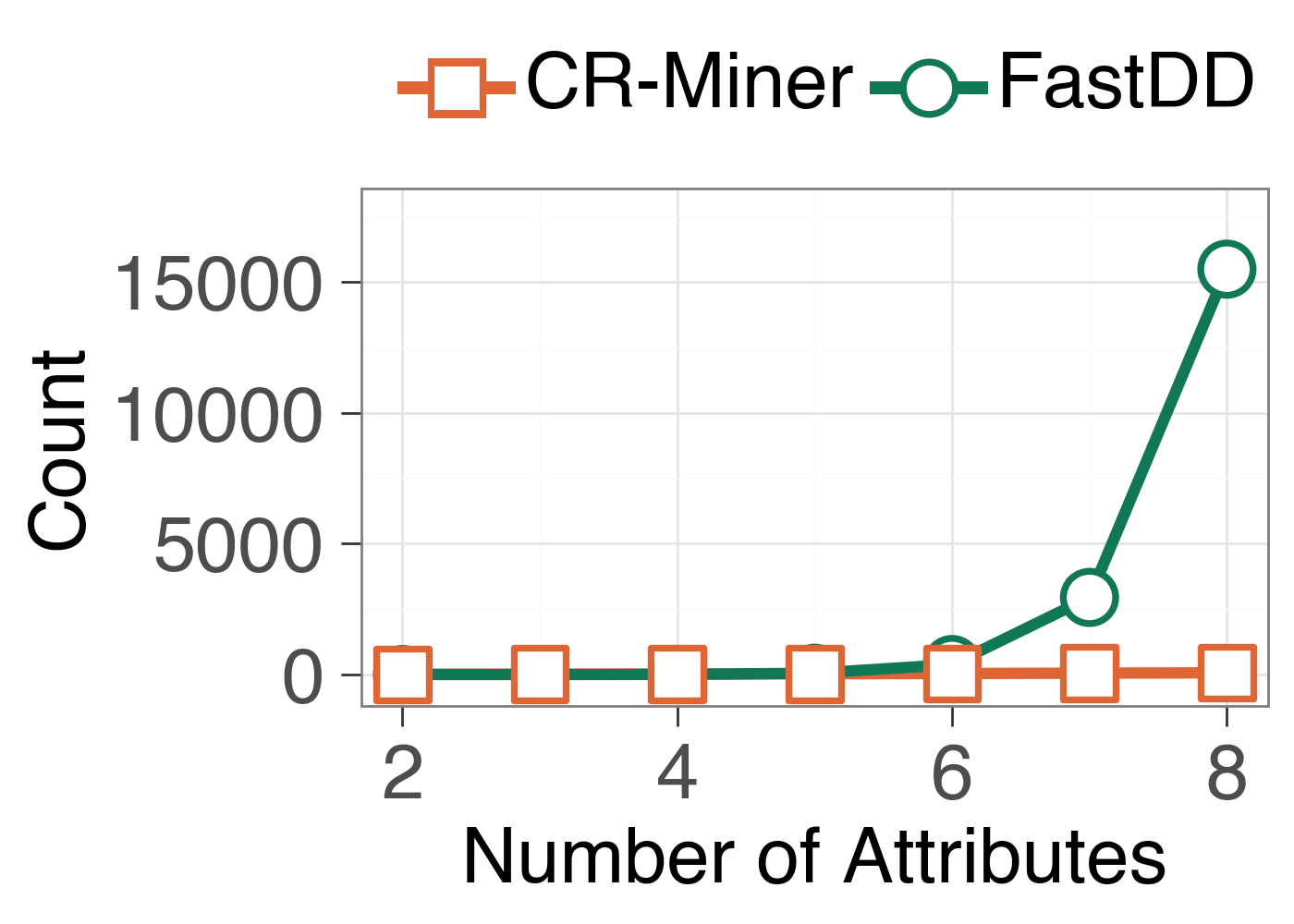}
        \caption{Power}
        \label{fig:powerexp4}
    \end{subfigure}
    \caption{Comparative number of discovered rules}
    \label{fig:depdatasize}
\end{figure}
% \vspace{-0.2cm}

\noindent \textbf{Exp-2: Runtime vs. \#attributes.} 
Figure~\ref{fig:overall_runtime} (e)-(h) shows the runtime \eat{number of discovered rules} as the number of attributes increases (data size is fixed at 20K). Across all datasets, \crminer consistently has a lower runtime compared to FastDD as the number of attributes increases. For MIMIC-III and Employment, \crminer exhibits stable, near-linear scaling, whereas FastDD runtimes increase at a faster rate, reflecting the increased computational overhead of pairwise value comparisons for more attributes. 

% \vspace{-0.2cm}

\noindent \textbf{Exp-3: \#Discovered rules vs. data size.}
Figure~\ref{fig:depdatasize} (a)-(c) shows the number of discovered CRs and DDs as the dataset size increases.  As expected, FastDD identifies a larger number of rules given that DDs subsume CRs, as the differential functions are expected to hold over all pairwise tuples in $I$.  In contrast, CRs hold over an ordered set of tuples. They are particularly important when sequential changes in both antecedent and consequent attributes are needed (among consecutive tuples) to identify causal relationships that are difficult to identify in DDs.  \crminer shows a stable set of frequent CRs for varying data sizes.

% \vspace{-0.2cm}

%FastDD produces a large number of dependencies across all datasets. The number of DDs generally decreases for both MIMIC-III and Weather as the data size grows, suggesting that the dependencies identified in a smaller set do not generalize and are pruned as more data is observed. Whereas for the Employment dataset, the no. of DDs grows with the data size. In contrast, \crminer yields a much smaller and more stable set of rules for all datasets. It remains consistently low, as only frequent and well-supported CRs are identified.

%On the Weather dataset, there is a significant runtime jump when moving from four to five attributes. This suggests that the fifth attribute introduces a more diverse set of differential functions, increasing the number of valid change rules. Despite this jump, \crminer remains more efficient.

% \begin{figure}[htbp]
%     \centering
    
%     \caption{Runtime comparison of DD and CR Discovery for increasing no. of attributes}
%     \label{fig:overall_runtimeAtt}
% \end{figure}

\noindent \textbf{Exp-4: \#Discovered rules vs. \#attributes.} Figure~\ref{fig:depdatasize} (d)-(f) shows the number of discovered CRs and DDs as the number of attributes increases (data size is fixed at 20K). As expected, FastDD shows an increased number of discovered rules across all datasets, due to an increased number of differential functions that must be evaluated, leading to more DDs.  We note that many of these DDs contain partial overlap among attributes and among the differential functions.  \crminer identifies non-redundant, minimal rules, leading to a more succinct and relevant set of rules.

% \vspace{-0.2cm}

%particularly in Employment and Weather, where the search space expands rapidly with higher number of attributes.  The increased number of differential   
%In contrast, \crminer maintains only slight increases and significantly lower count. 
%The number of discovered rules remains stable even as dimensionality grows due to the high support constraint ($\theta = 0.9$).

% \begin{figure}[htbp]
%     \centering
    
%     \caption{Number of of DDs and CRs discovered for increasing no. of attributes}
%     \label{fig:depAttr}
% \end{figure}

\begin{figure}[t]
    \centering
    \begin{subfigure}[b]{0.45\linewidth}
        \centering
        \includegraphics[width=0.7\linewidth]{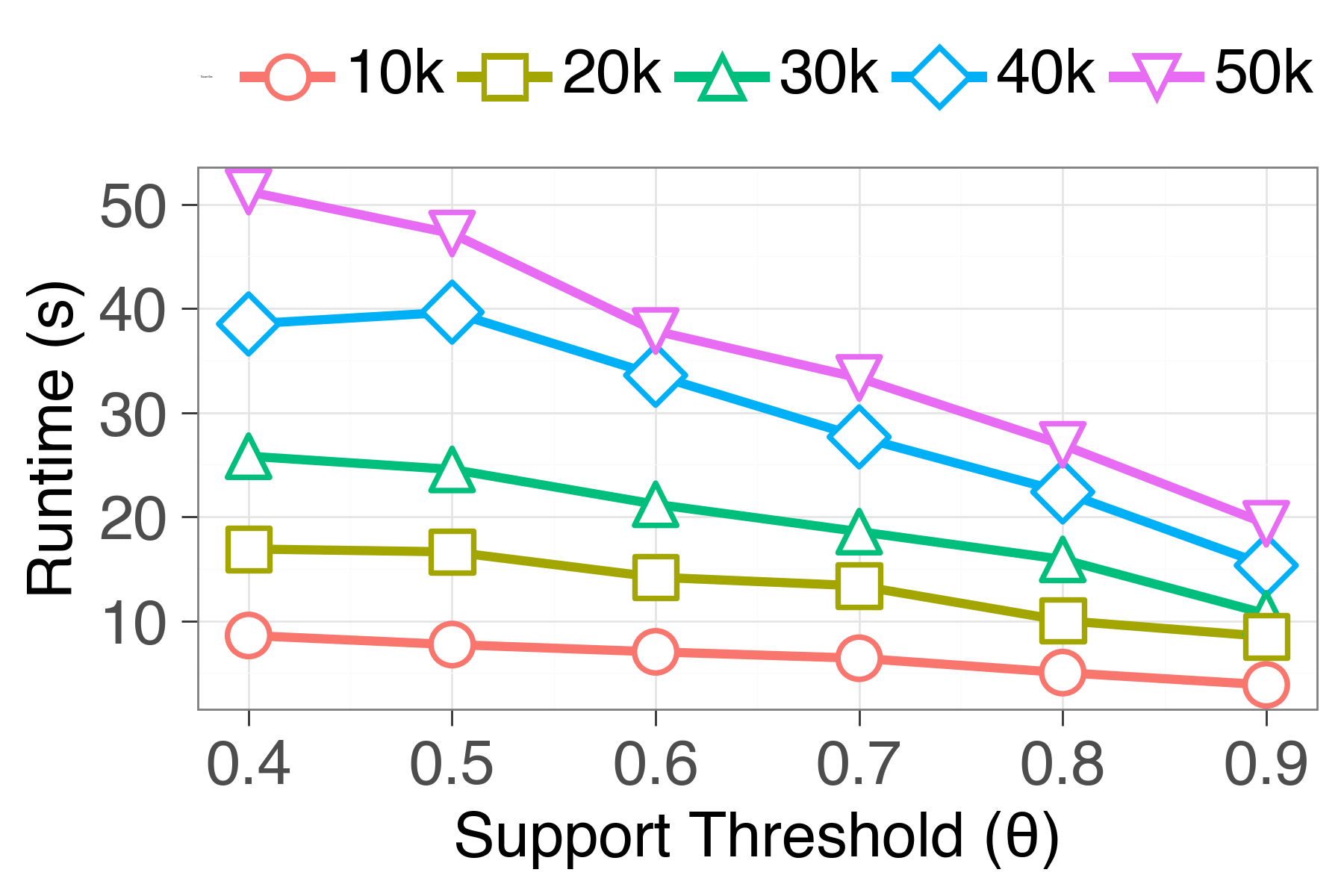}
        \caption{\crminer runtime}
        \label{fig:supprun}
    \end{subfigure}
    \hfill
    \begin{subfigure}[b]{0.45\linewidth}
        \centering
        \includegraphics[width=0.7\linewidth]{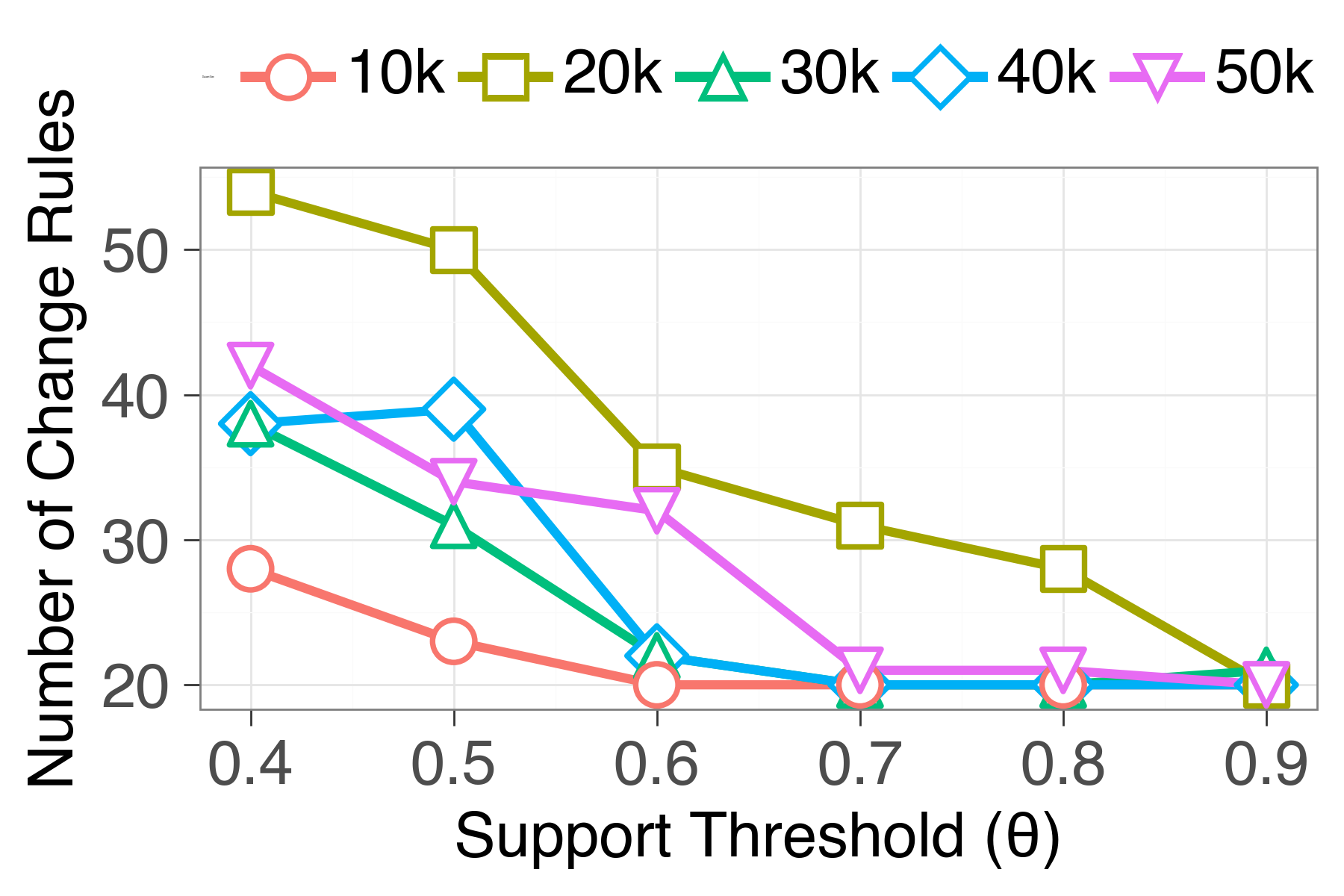}
        \caption{No. of CRs}
        \label{fig:suppcr}
    \end{subfigure}
    \caption{Varying support threshold}
    % \vspace{-0.2cm}
\end{figure}

\noindent \textbf{Exp-5: Varying support.} Figure~\ref{fig:supprun} shows the \crminer runtime for varying support, and data sizes using the MIMIC-III dataset.  For all data sizes, the runtimes linearly decrease as support levels increase, as expected.  Higher support thresholds enable more aggressive pruning of the search space, as fewer CR candidates satisfy the stricter criteria. \eat{This downward trend is more pronounced in larger datasets (e.g., 50k tuples), where the reduction in candidate evaluation significantly offsets the overhead for a larger number of tuples.}  Consequently, Figure~\ref{fig:suppcr} shows the number of discovered CRs for varying support. As the support threshold increases, the number of discovered CRs decreases, as expected.

\begin{figure}[t]
    \centering
    \begin{subfigure}[b]{0.3\linewidth}
        \centering
        \includegraphics[width=\linewidth]{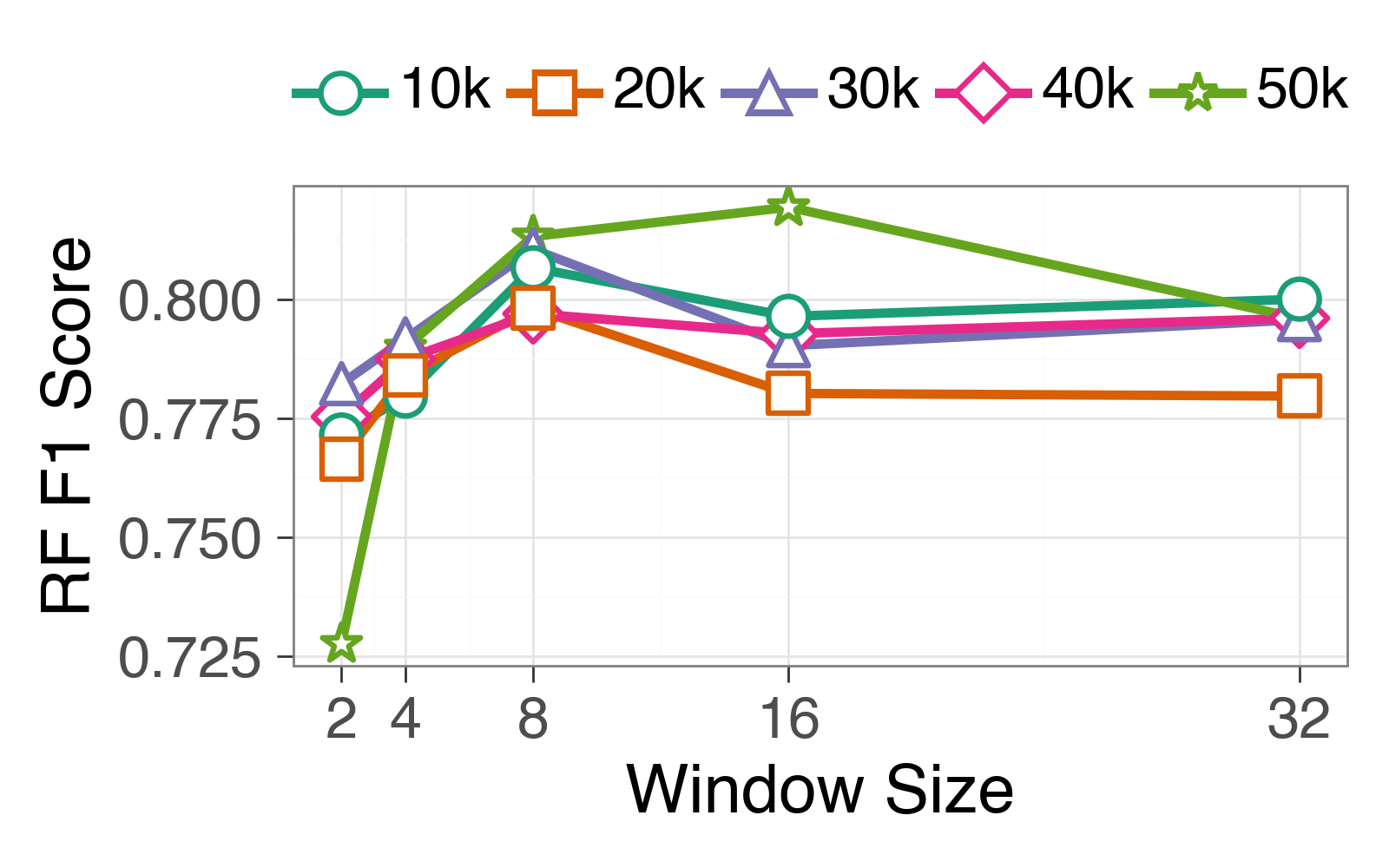}
        \caption{F1 Score for changing $\mathcal{W}$}
        \label{fig:f1}
    \end{subfigure}
    \hfill
    % \begin{subfigure}[b]{0.3\linewidth}
    %     \centering
    %     \includegraphics[width=\linewidth]{Images/7b.png}
    %     \caption{Recall}
    %     \label{fig:recall}
    % \end{subfigure}
    % \hfill
    % \begin{subfigure}[b]{0.3\linewidth}
    %     \centering
    %     \includegraphics[width=\linewidth]{Images/7c.png}
    %     \caption{Precision}
    %     \label{fig:prec}
    % \end{subfigure}
    % \hfill
    % \begin{subfigure}[b]{0.3\linewidth}
    %     \centering
    %     \includegraphics[width=\linewidth]{Images/7d.png}
    %     \caption{Target CRs found}
    %     \label{fig:targetCRs}
    % \end{subfigure}
    % \hfill
    % \begin{subfigure}[b]{0.3\linewidth}
    %     \centering
    %     \includegraphics[width=\linewidth]{Images/7e.png}
    %     \caption{Total Runtime}
    %     \label{fig:runtimewindow}
    % \end{subfigure}
    % \hfill
    \begin{subfigure}[b]{0.3\linewidth}
        \centering
        \includegraphics[width=\linewidth]{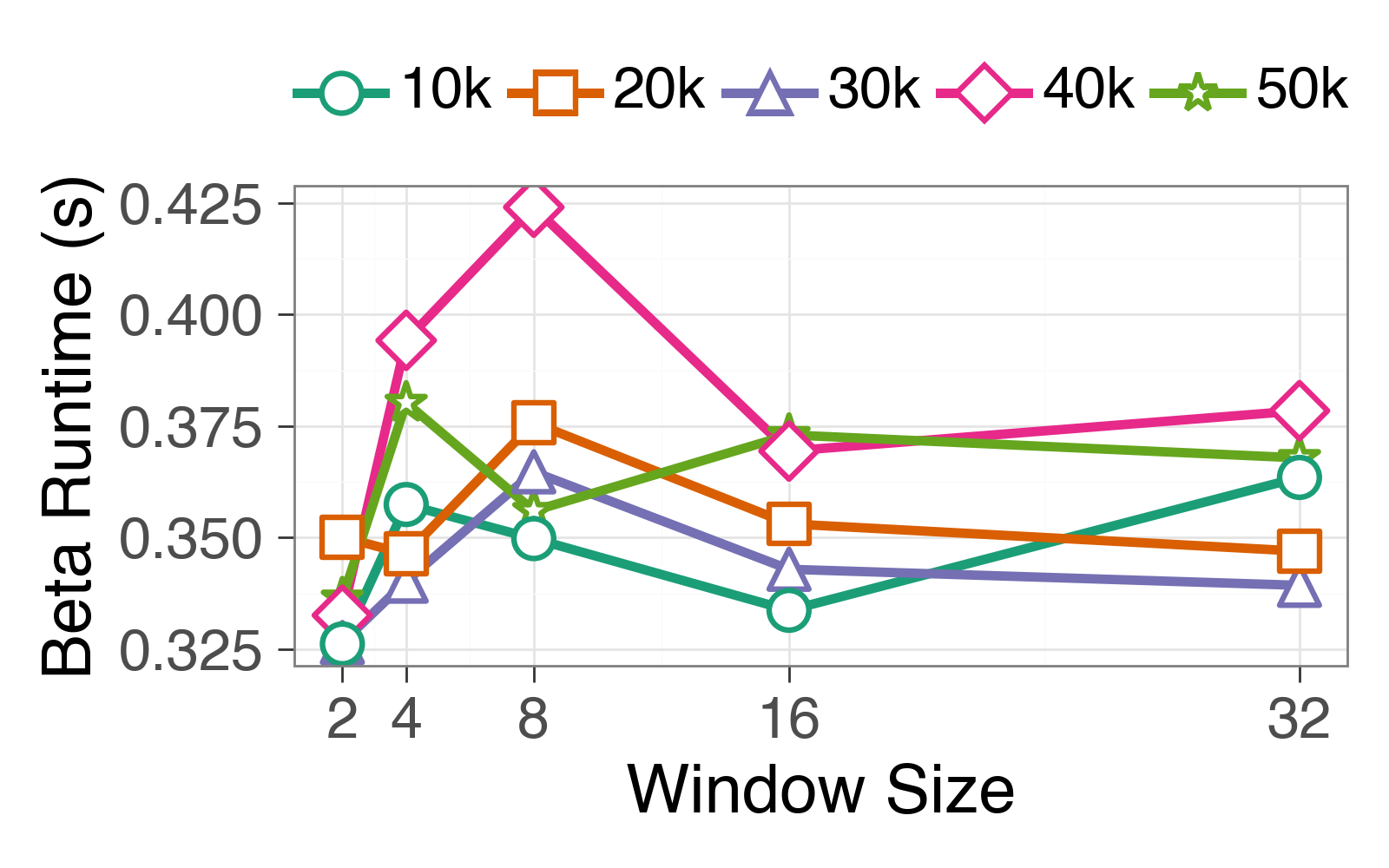}
        \caption{Beta Runtime for changing $\mathcal{W}$}
        \label{fig:betaruntime}
    \end{subfigure}
    \hfill
    \begin{subfigure}[b]{0.3\linewidth}
        \centering
        \includegraphics[width=\linewidth]{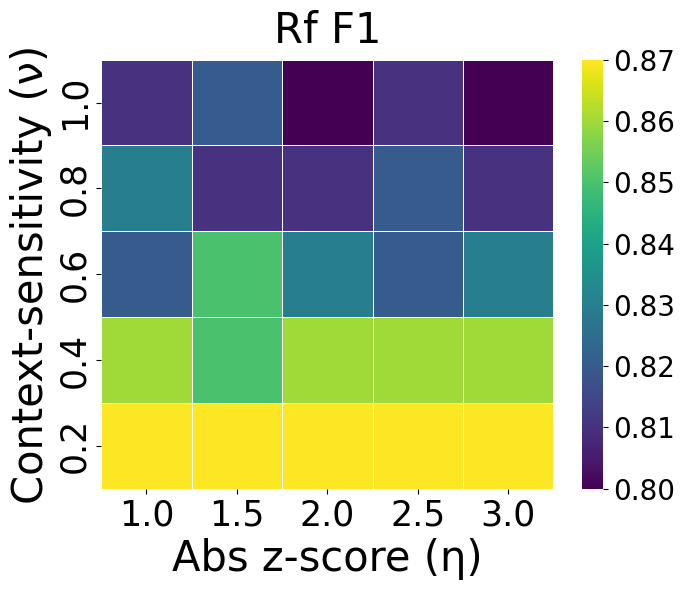}
        \caption{F1 Score for changing $\eta$ and $\nu$}
        \label{fig:f1label}
    \end{subfigure}

    \caption{\crminer Performance}
    \label{fig:perf}
\end{figure}

\noindent \textbf{Exp-6: Varying window size $\mathcal{W}$.} Figures~\ref{fig:perf} (a)-(b) show the impact of varying the window size across different dataset sizes (10k to 50k) using the MIMIC-III dataset. As shown in Figure~\ref{fig:f1}, the Random Forest F1 Score peaks for most dataset configurations when the window size is between 8 and 16. For the 50k dataset, a sharp dip in F1 score is observed at the smallest window size of 2, indicating that overly restricted windows might introduce local noise or capture insufficient context for stable z-score normalization. 

% \vspace{-0.1cm}

% Figure~\ref{fig:window} (b) illustrates the Target CRs found across varying window sizes. While most dataset scales remain relatively stable within a tight range of 8 to 13 rules, the 20k dataset shows an upward trajectory, peaking at 20 rules when the window size reaches 32. This behaviour occurs because larger window sizes expand the historical context considered for each data point, allowing the algorithm to smooth out short-term fluctuations and unearth change rules that remain hidden within smaller, more localized windows. 

% \vspace{-0.1cm}
Figure~\ref{fig:betaruntime} displays the runtime for calculating the context-score $\beta$. The execution time for $\beta$ calculation peaks when the window size is between 4 and 8, but then drops and remains almost consistently low. This aligns with the fact that larger window sizes result in fewer identified abnormal change labels, which ultimately leads to a shorter computation time for $\beta$. 

% the window size does not introduce severe computational overhead, making the system highly robust to parameter adjustments.

% \vspace{-0.2cm}
% Overall, \crminer demonstrates robust performance across the evaluated window sizes, with F1 scores, recall, and precision remaining consistently in the 0.78--0.82 range for most dataset sizes once $\mathcal{W} \ge 8$. The most notable sensitivity appears at the smallest window size ($\mathcal{W} = 2$), where the 50k dataset exhibits a pronounced dip in F1 and recall before recovering sharply at $\mathcal{W} = 4-8$. This suggests that very small windows produce overly noisy local baselines, causing instability in z-score normalization. As the window size increases beyond 8, performance stabilizes comparatively, indicating that \crminer is not highly sensitive to the choice of $\mathcal{W}$ in the moderate-to-large range.

% Figure~\ref{fig:window} (d) shows that the number of target CRs discovered remains broadly stable across window sizes for most dataset configurations, with the exception of the 20k dataset, which exhibits a notable increase in discovered CRs around $\mathcal{W} = 16$. {\color{blue}but why?} Runtime metrics in Figure~\ref{fig:window} (e)--(f) confirm that varying $\mathcal{W}$ has a negligible impact on both total and beta runtime, demonstrating that \crminer's computational cost is not sensitive to window size.

\noindent \textbf{Exp-7: Varying absolute z-score threshold $\eta$ and context-sensitivity threshold $\nu$.} Figure~\ref{fig:f1label} shows F1 score as a function of the absolute z-score threshold $\eta$ and context-sensitivity parameter $\nu$. The highest F1 scores are concentrated along the bottom rows of the heatmap (low $\nu$, 0.2--0.4) across all values of $\eta$, indicating that the random forest model performs best when abnormality labelling is driven primarily by the absolute z-score criterion rather than contextual deviation. Performance degrades progressively as $\nu$ increases, suggesting that over-weighting the context-relative condition introduces noise into the labelling process.

% \begin{figure}[t]
%     \centering
%     \begin{subfigure}[b]{0.3\linewidth}
%         \centering
%         \includegraphics[width=\linewidth]{Images/8a.png}
%         \caption{F1 Score}
%         \label{fig:f1}
%     \end{subfigure}
%     \hfill
%     \begin{subfigure}[b]{0.3\linewidth}
%         \centering
%         \includegraphics[width=\linewidth]{Images/8b.png}
%         \caption{Recall}
%         \label{fig:recall}
%     \end{subfigure}
%     \hfill
%     \begin{subfigure}[b]{0.3\linewidth}
%         \centering
%         \includegraphics[width=\linewidth]{Images/8c.png}
%         \caption{Precision}
%         \label{fig:prec}
%     \end{subfigure}
%     % \hfill
%     % \begin{subfigure}[b]{0.3\linewidth}
%     %     \centering
%     %     \includegraphics[width=\linewidth]{Images/8d.png}
%     %     \caption{Target CRs found}
%     %     \label{fig:targetCRs}
%     % \end{subfigure}
%     % \hfill
%     % \begin{subfigure}[b]{0.3\linewidth}
%     %     \centering
%     %     \includegraphics[width=\linewidth]{Images/8e.png}
%     %     \caption{Total Runtime}
%     %     \label{fig:runtimewindow}
%     % \end{subfigure}
%     % \hfill
%     % \begin{subfigure}[b]{0.3\linewidth}
%     %     \centering
%     %     \includegraphics[width=\linewidth]{Images/8f.png}
%     %     \caption{Beta Runtime}
%     %     \label{fig:betaruntime}
%     % \end{subfigure}
%     \caption{Studying the effects of changing labelling thresholds}
%     \label{fig:labelling}
% \end{figure}

% \noindent \textbf{Exp-9: Varying support variation threshold $\epsilon_s$.}
% \vspace{-0.2cm}

\begin{figure}[t]
    \centering
        \includegraphics[width=0.6\linewidth]{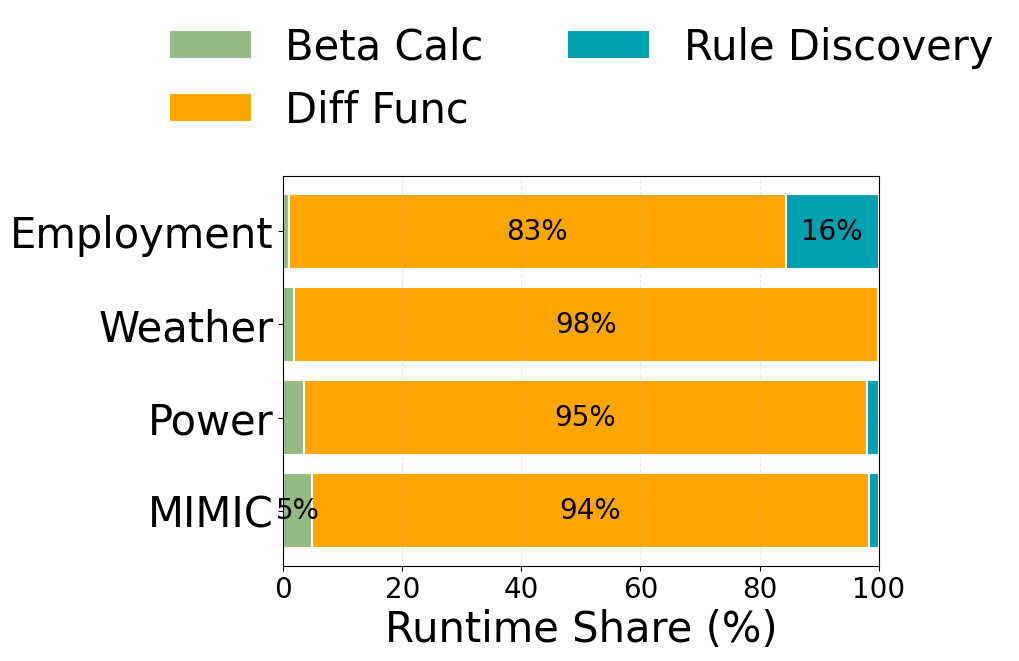}
    \caption{Runtime for each module of the framework}
    \label{fig:ablation}
\end{figure}
\noindent \textbf{Exp-8: Runtime Breakdown.} Figure~\ref{fig:ablation} illustrates the runtime breakdown of each major module of the \crminer framework across different datasets. Across all four datasets, differential function discovery consistently accounts for the majority of the total execution time. In contrast, context-factor $\beta$ calculation takes a negligible fraction of the runtime across the board. The contribution of rule discovery varies by dataset. Rule discovery takes longer for the employment dataset because several attributes are large-scale aggregate values with drastic changes between consecutive tuples. This produces very large change intervals that are frequently satisfied by many tuple pairs, which makes antecedent–consequent intersection computation more expensive. 

% It constitutes a significant portion of the execution time for the Employee dataset compared to the others.

% Consequently, the total runtime is primarily driven by the efficiency of the differential function discovery step, as expected.

% \vspace{-0.7cm}

\noindent \textbf{Exp-9: Example Change Rules.} We present three examples of CRs mined from the Weather, MIMIC, and Power data.\\
\noindent \textbf{$\phi_1$:} [MinTemp] $_{(0, 0.2)}\rightarrow_{(0, 0.1)}$ [CoolDegDays]. MinTemp captures the daily minimum temperature, while CoolDegDays measures cooling energy demand. Small changes in MinTemp do not contribute to higher daytime temperatures and therefore do not increase cooling demand. Consider the DD \\$\upsilon_1:{\text{[MinTemp}(>12.6)] \land \text{[MeanTemp}(\le 0)] \rightarrow \text{[CoolDegDays}(>1.3)]}$, which is defined over unordered tuple pairs.  Unfortunately, $\upsilon_1$ is unable to characterize how consecutive changes in the MinTemp trigger changes in the cooling demand. \\
\noindent \textbf{$\phi_2$:} [Hct] $_{(0, 0.2)}\rightarrow_{(-2.7, 2.7)}$ [Hbg]. $\phi_2$ reflects the well-established relationship between Hematocrit (Hct) and Hemoglobin (Hbg), commonly approximated by the clinical "The Rule of Three," where Hct $\approx 3 \times$ Hbg \cite{celkan2020does}. The CR $\phi_2$ exemplifies this constraint, showing a restricted change of Hbg up to 3x, reflecting a clinical biological constraint. In contrast, the DD $\upsilon_2: \{[\text{Hct}(>29.7)] \land [\text{RBC}(\le 0)] \rightarrow $$[\text{Hbg}(>10.1)]\}$  defines differences in RBC and Hematocrit that trigger Hemoglobin changes \emph{among non-consecutive} records, making sequential trend analysis not possible.\\
\noindent \textbf{$\phi_3$:} [Humidity] $_{(-6.7, 96.8)} \rightarrow_{(-9211.14, 5711.34)}$ [Zone1]. $\phi_3$ models variations in humidity corresponding to substantial changes in zone energy demand. This reflects the sensitivity of HVAC-driven systems to atmospheric conditions, where humidity variations influence cooling and dehumidification requirements, thereby affecting overall power consumption. In contrast, $\upsilon_3: \{[\text{Temperature}(\le 0.2)] \land [\text{Humidity}(\le 0.0)] \land [\text{gendiffuseflows}(\le 0.01)] \land [\text{Zone3}(\le 0.0)] \rightarrow [\text{Zone1}(>6000.0)]\,\}$ only captures a conjunction of conditions leading to high Zone 1 consumption, without modeling how changes evolve across consecutive observations. 

% \section{Limitations}
%     \input{Chapters/Limitations}

\section{Related Work}
    \noindent \textbf{Exploring Change.} Prior work on change exploration focuses on detecting and interpreting data evolution. Early approaches, such as \cite{chawathe1997meaningful}, model the semantics of individual changes but do not capture interactions across attributes. The "change cube" framework \cite{bleifuss2018exploring} enables multi-dimensional analysis of temporal changes, but remains limited to exploration and does not model inter-attribute propagation effects. Clustering-based approaches for time-series data \cite{bornemann2018data} identify common temporal patterns, but treat entity-attribute pairs independently, ignoring dependencies across attributes. Similarly, streaming-based frameworks \cite{kanza2023data} detect changes using statistical signals, but focus only on anomaly detection.

\noindent \textbf{Data Dependencies.} Data dependencies capture relationships between attributes. Order Dependencies (ODs) enforce ordering relationships but do not quantify attribute-level changes \cite{szlichta2016effective}. Sequential Dependencies (SDs) bound the changes in the consequent for ordered tuples, and Conditional Sequential Dependencies (CSDs) extend this with interval-based validity and support measures \cite{golab2009sequential}. However, they only focus on changes in the consequent. Differential Dependencies (DDs) model bounded differences in both antecedent and consequent attributes across tuple pairs \cite{song2011differential}. While expressive, they ignore tuple ordering and do not capture sequential patterns. Conditional Differential Dependencies (CDDs) extend DDs by restricting dependencies to subsets of tuples defined by condition sets \cite{kwashie2015conditional}. However, CDDs also operate on unordered tuple pairs and do not capture the context of a change.

\noindent \textbf{Statistical and ML Approaches.} Statistical and machine learning methods detect abnormal changes by modeling expected behaviour and identifying deviations. Statistical techniques, such as Z-scores \cite{heckert2002handbook}, model expected patterns of behaviour through assumptions about distributions. For temporal data, control charts and statistical process control (SPC) are employed to identify outliers that exceed predefined control limits \cite{qiu2017statistical}. These methods are only effective when the data conforms to known distributions. Learning approaches such as isolation-based methods \cite{liu2008isolation}, deep autoencoder-based models \cite{sakurada2014anomaly}, and sequence models like LSTMs \cite{malhotra2015long} can capture complex patterns in data and detect deviations from learned behaviour. However, these methods focus on anomaly detection and do not provide interpretable relationships between attribute changes.

\section{Conclusion and Future Work}
    We introduce \emph{change rules (CRs)}, which extend existing data quality rules by defining bounds on allowed changes in both the antecedent $X$ and consequent $Y$ attributes for tuples ordered on $X$.  Our discovery framework identifies the relevant context for attribute changes and applies a scaling factor to adjust for out-of-norm changes.  We introduce \crminer, a mining algorithm that generates and identifies minimal intervals leading to \eat{frequent }CRs satisfying a given support threshold.  Our experimental evaluation shows that \crminer is scalable for increasing data sizes and number of attributes, with a 40-50\% runtime improvement over existing differential dependency mining baselines. As next steps, we plan to first extend $X$ to multiple antecedent attributes to capture joint attribute relationships and increase expressiveness.  Secondly, we will study semantic extensions to consider conditional constraints where CRs hold, e.g., $\phi'_{2}:$ [\emph{Province = Ontario}, MinTemp] $_{(0, 0.2)}\rightarrow_{(-0.1, 0)}$ [CoolDegDays], states that $\phi'_{2}$ holds only for records within the province of Ontario.

\bibliographystyle{splncs04}
\bibliography{biblio}
\end{document}